%% file: GP_final.tex
\documentclass[manuscript]{aastex}

\usepackage{natbib}
\usepackage[latin1]{inputenc}
\usepackage[T1]{fontenc}
\usepackage[english]{babel}
\usepackage{amsfonts}
\usepackage{eucal}
\usepackage{amsthm}
\usepackage{amsbsy}
\usepackage{amssymb}
\usepackage{amsmath}
\usepackage{wasysym}
\usepackage{graphicx}
\usepackage{babel}
\usepackage{fancyhdr}
\usepackage{float}
\usepackage{multicol}
\usepackage{textcomp}
\usepackage[caption=false]{subfig}
\usepackage{epsfig}
\usepackage{ mathrsfs }
\usepackage{verbatim}

\usepackage{url}

\renewcommand{\vec}[1]{\mathbf{#1}}

\begin{document}

%% LaTeX will automatically break titles if they run longer than
%% one line. However, you may use \\ to force a line break if
%% you desire.

\title{Gaussian Process for star and planet characterisation}

%% Use \author, \affil, and the \and command to format
%% author and affiliation information.
%% Note that \email has replaced the old \authoremail command
%% from AASTeX v4.0. You can use \email to mark an email address
%% anywhere in the paper, not just in the front matter.
%% As in the title, use \\ to force line breaks.

\author{C. Danielski}
\affil{Dept. of Physics \& Astronomy, University College London, Gower Street, WC1E 6BT, UK\\
Institut d'astrophysique spatiale, Universit\'{e} Paris-Sud 11 $\&$ CNRS (UMR 8617)}
\email{camilla.danielski@ias.u-psud.fr}
\author{T. Kacprzak, G. Tinetti, P. Jagoda}
\affil{Dept. of Physics \& Astronomy, University College London, Gower Street, WC1E 6BT, UK}

%% Mark off your abstract in the ``abstract'' environment. In the manuscript
%% style, abstract will output a Received/Accepted line after the
%% title and affiliation information. No date will appear since the author
%% does not have this information. The dates will be filled in by the
%% editorial office after submission.

\begin{abstract}
 The study of exoplanetary atmospheres epitomises a continuous quest for higher accuracy measurements. 
Systematic effects and noise associated with both the stellar activity and the instrument can bias the results and thus limit the precision of the analysis.
To reach a high photometric and spectroscopic precision, it is therefore essential to correct for these effects.
We present here a novel non-parametric approach, named \textit{Gaussian Process method for Star Characterization} (GPSC), to remove effects of stellar activity and instrumental systematics on planetary signals, with a view to preserve the atmospheric contribution which can be as small as 10$^{-4}$ or even 10$^{-5}$ the flux of the star.\\
We applied our method to data recorded with Kepler, focussing on a sample of lightcurves with different effective temperatures and flux modulations. We found that GPSC can very effectively correct for the short and long term stellar activity and instrumental systematics.
Additionally we run the GPSC on both real and simulated transit data, finding transit depths consistent with the original ones.
Consequently we considered 10-hours of continuous observations: daily, every other day and weekly, and we used the GPSC to reconstruct the lightcurves.
When data are recorded more frequently than once every five days we found that our approach is able to extrapolate the stellar flux at the 10$^{-4}$ level compared to the full stellar flux.

These results show a great potential of GPSC to isolate the relevant astrophysical signal and achieve the precision needed for the correction of short and long term stellar activity.
\end{abstract}

%% Keywords should appear after the \end{abstract} command. The uncommented
%% example has been keyed in ApJ style. See the instructions to authors
%% for the journal to which you are submitting your paper to determine
%% what keyword punctuation is appropriate.

%% From the front matter, we move on to the body of the paper.
%% In the first two sections, notice the use of the natbib \citep
%% and \citet commands to identify citations.  The citations are
%% tied to the reference list via symbolic KEYs. The KEY corresponds
%% to the KEY in the \bibitem in the reference list below. We have
%% chosen the first three characters of the first author's name plus
%% the last two numeral of the year of publication as our KEY for
%% each reference.

%% Authors who wish to have the most important objects in their paper
%% linked in the electronic edition to a data center may do so by tagging
%% their objects with \objectname{} or \object{}.  Each macro takes the
%% object name as its required argument. The optional, square-bracket 
%% argument should be used in cases where the data center identification
%% differs from what is to be printed in the paper.  The text appearing 
%% in curly braces is what will appear in print in the published paper. 
%% If the object name is recognized by the data centers, it will be linked
%% in the electronic edition to the object data available at the data centers  
%%
%% Note that for sources with brackets in their names, e.g. [WEG2004] 14h-090,
%% the brackets must be escaped with backslashes when used in the first
%% square-bracket argument, for instance, \object[\[WEG2004\] 14h-090]{90}).
%%  Otherwise, LaTeX will issue an error. 

\section{Introduction}
In the study of exoplanetary atmospheres, stellar activity plays a critical role. 
Previous analyses have shown how the intrinsic variation of a host-star can affect the estimate of the planetary system parameters, 
both for radial velocity and transit measurements (e.g. Czesla et al. 2009, Lagrange et al. 2010, Boisse et al. 2011, Ballerini et al. 2012, Oshagh et al. 2013).
It is therefore important to identify and remove the photospheric stellar activity in order to recover the uncontaminated planetary signal,
especially in a condition where planetary atmospheric features in the infrared have the contrast of 10$^{-4}$ with respect the host-star radiation.
This problem can be tackled in two different ways: either by studying the photometric modulations of the stellar flux as a function of time 
or by using stellar spectral information in the visible band to correct the IR spectrum.
However, for the second approach, there are no instruments available yet that provide a simultaneous coverage over such wide spectral range.\\
Here we focused on the photometric approach applied to data recorded from space.\\
Among the available instruments, the NASA Kepler mission (Borucki et al 2010, 2011) provided very accurate observations of thousands of stars.
With its unprecedented photometric precision, Kepler is capable of groundbreaking discoveries: 
the detection of the moon-sized planet Kepler-37b (Barclay et al. 2013), 
the finding of the two super-Earth sized planets in the habitable zone around Kepler-62 (Borucki et al. 2013)
or the discovery of the first Earth-size planet with measured Earth-mass, Kepler-78b (Sanchis-Ojeda et al 2013, Howard et al. 2013, Pepe et al. 2013).
These discoveries suggest that small worlds are common and, most importantly, are observable today if the host star is bright enough, and if we reach the  
sufficient photometric precision. 
To detect the smallest flux variations, the data reduction techniques adopted are critical: instrumental systematics and stochastic errors, in particular, need meticulous corrections (Jenkins 2010a).
Intense work in this regard has been done by the Kepler team: the Presearch Data Conditioning (PDC) module of the Kepler data analysis pipeline (Jenkins et al. 2010b, Gilliland et al. 2010b, Christiansen et al. 2012) is very efficient at extracting transit signals. 
However, it was not optimized to analyse the low-frequency stellar signal (Murphy 2012) and some instrument systematics may still appear in the residuals after the correction.
Alternative ways to address these issues have been suggested: for example Garc\'{i}a et al. (2011) discussed the process of correction of Kepler time-series focusing more on asteroseismic applications. 
Smith et al. (2012) applied a Bayesian Maximum A Posteriori approach to model the instrumental systematics by correlating multiple non-active targets.
All these approaches either apply parametric corrections, which potentially may distort the signal and inject supplementary noise, or use a limited model class (e.g. linear), that may not catch the real underlying stellar trend. 
The problem can be mitigated by adopting a non-parametric approach, which makes no assumptions about the functional form of the underlying stellar brightness function or its parameters.
Instead it uses assumptions about the statistical properties of the function, and infers it from the data itself.

Non-parametric techniques have been used in a variety of astronomical contexts (e.g. Seikel et al. 2012, Chapman et al. 2013). 
In the exoplanet field in particular, these methods have been adopted by e.g. Thatte et al. 2010, Gibson et al. 2012a, Ford et al. 2012, Waldmann 2012, 2013, Waldmann et al. 2013a.  
Among these the Gaussian Process (GP) method for regression (Rasmussen $\&$ Williams, 2006) is widely used in the machine learning field and of increasing interest in the exoplanet community. 
GPs are defined as \textquotedblleft non-parametric\textquotedblright ~as no parametric function is used to model the regression function.
The model contains a covariance function and its parameters, but this function describes the statistical properties of the regression function, rather than the regression function itself.

We developed a technique, named \textit{Gaussian Process method for Star Characterization} (GPSC), to analyse the stellar flux modulations.
The method uses a Gaussian Process to reconstruct these modulations from the data, precisely accounting for measurement noise, gaps in observations, planetary transits, flares and other sudden variations in brightness.
The core of this technique is the precise modelling of the kernel function of a GP, which controls the lengths of the stellar modulations, so that both long and short scale trends are accounted for in great detail.
We find that GPSC can effectively model the stellar luminosity function to a degree that can be successfully for characterisation of planetary atmospheres.

We demonstrated the efficiency of the technique by applying it on Kepler lightcurves, chosen because of the continuity of the observations and the above-mentioned photometric precision.
We expect a similar efficiency when applied to a wider range of similar datasets.
We used GPSC on a sample of ten stars with different effective temperatures. 
We de-trended the short and long term stellar activity and extracted the temporal \textit{events} intrinsic to the stellar system, such as planetary transit or flares. 
Notice that we call \textquotedblleft short\textquotedblright ~and \textquotedblleft long\textquotedblright
~term respectively the high frequencies and low frequencies modulations on a scale of $\sim$90 days i.e. the average duration of a Kepler quarter.  
A quarter is a terminology used to define each season of Kepler data collection.
The de-trended residuals between the data and the model show no visible auto-correlation or structure, indicating good performance of the method.
We run the GPSC on both real and simulated data to test its de-trending ability, finding transit depths consistent with the original ones.

With the same approach we also extrapolated the stellar flux in a simulated periodic monitoring scenario, recovering the original stellar flux with a 10$^{-4}$ accuracy.

\section{Method}

\subsection{\textit{An introduction to Gaussian Processes}}
 \label{GP}
\input{section_gp.tex}

\subsubsection{\textit{Statistical model} }\label{sec:model}
We describe here the specifications of our statistical model for the observed lightcurves. 
We also state the assumptions made to model the stellar flux modulations and the events (\S\ref{sec:data_analysis}).

\begin{enumerate}
	\item The true luminosity of a star can be estimated through a covariance function, which is of the form expressed in in Eqn (\ref{eqn:rbf_kernel}).
	The true physical process which governs the luminosity of the star is assumed unknown. 
	We also assume that a combination of RBF kernels with different widths is a good model for the luminosity variation.
	In this work we found that a kernel with two RBF components works well, and using the Occam's Razor principle we did not use more complicated models.
	To find the hyperparameters of the RBF kernel, we maximise the marginal likelihood of the data using the conjugate gradients algorithm.
	This routine is available with GPML package (see GPML documentation\footnote{ http://www.gaussianprocess.org/gpml/code/matlab/doc/} for details). Marginal likelihood is a non-convex function of the kernel parameters and thus can have local maxima. 
	To find the global maximum, we start the minimizer 20 times, initialized with random kernel parameters.
	This procedure does not guarantee to find the global maximum, however, we found that it is sufficient to find a satisfactory solution.

	\item The observed values of stellar luminosity are contaminated with additive Gaussian noise with some standard deviation, which can be roughly estimated from the data. 
	Although this is an important parameter, we find that a small error in its estimation does not change the results significantly.

	\item Except for Gaussian noise, the instrument may introduce a systematic variation in the observed stellar luminosity, for example a periodic oscillation (see \S\ref{sec:data_analysis}).
	If this effect is present we do not treat it separately from the luminosity of a star and instead use a GP to model both simultaneously.
	
	\item \label{time_event} The raw data show events which are not governed by the same physical process directly responsible for the stellar variation of the luminosity.  
	We identified as events the data points whose luminosity lay beyond the $3\sigma$ threshold of the GP probability distribution on the luminosity function (Eqn \ref{eqn:gp_conditional}). 
	We specified a parameter $t^{event}_{max}$ [days], as being the maximum duration of an event expected in the data. 
	We removed these data points and a range of neighbouring points, which lay within $t^{event}_{max}$ range of the identified event points. 
	Consequently we re-trained the GP to model the stellar luminosity without the influence of these outliers 
	(see \S\ref{sec:data_analysis}, Fig. \ref{raw_inliers}).

	\item \label{time} We used a single GP to model each time-series. 
	GP inference requires an inverse of a matrix of size $N^{2}$, where $N$ is the number of training points, and is a $\mathcal{O}(N^3)$ process. 
	On a standard desktop computer, this is feasible in reasonable time for data sets with less than 10\,000 points. 
	Data sets we analysed had fewer points.
	However, if more points are needed in the time-series, there exist large scale implementations of GP (see Qui$\~{n}$onero-Candela 2005 for reviews).

\end{enumerate}

\subsection{\textit{Data analysis}}
\label{sec:data_analysis}
The purpose of this paper is twofold.
First we want find astrophysical events, such as flares, planetary transits, occultations or general sudden variations in the intensity. 
Second we want to be able to predict the stellar flux at the 10$^{-4}$ level in a situation where a constant monitoring of the star is not available. 

We investigated a selection of stars belonging to the Kepler Objects of Interest (KOI) catalog, with different effective temperatures $T_{eff}$, i.e. with
different photometric variability (Basri et al. 2010, 2013).
For each star in the sample we analysed the long cadence (LC) Simple Aperture Photometry (SAP) data of all the available quarters (Table \ref{targets}).
All the time-series we studied exhibit data-breaks, discontinuities, flux jumps and drifts (Garc\'{i}a et al. 2011).  
The data-breaks are generally due to the monthly downloads or to the space telescope safe mode and loss-of-pointing. 
The safe mode is normally responsible for the longest gaps in the data, 
followed by a sudden exponential increase/decrease in flux as the telescope returns to its nominal temperature and focus. 
Unexpected jumps in flux can also be due to pointing tweaks and pixel sensitivity drops (Jenkins et al. 2010a).  
These instrumental effects are in part corrected in the PDC flux, however it has been shown that in some cases PDC-flux data may modify the astrophysical signals 
(i.e. low-frequency stellar signals such as the ones produced by long/short lived star-spots) not directly related to exoplanetary transits (Garc\'{i}a et al 2011, Murphy 2012). \\
Furthermore the collection of the long cadence (LC) data does not require the hardware to do anything different than what it does for the short-cadence (SC) data.
Hence, the level of counts as a function of time is equivalent for both LC and SC data.
The only difference between them is the integration time: the LC pixels, co-added in the Science-data-accumulator, are read once every 29.4 min (270 exposures) while the SC are read every 58.9 s (9 exposures).
For these reasons we chose to use for this study the SAP-flux data (henceforth referred to us the {\textquotedblleft raw\textquotedblright ~data), analysing each quarter separately.

Before proceeding with our analysis we normalized each time-series of the stellar sample. 
Then we removed on average $\sim$50-100 points, equivalent to $\sim$1.04-2.08 days,  around systematic gaps longer than 0.5 BJD (Barycentric Julian date). 
Our choice of cutting the data at the extremities of these gaps does not affect our analysis and it is a luxury we can afford because of the very large number of data points in the dataset.  
This conservative approach prevents the addition of further systematic trends to our GPSC model. 
The number of data points to cut was tailored to every single lightcurve.

\subsubsection{Detrending of the long-term stellar activity}\label{sec:detrending}

The typical molecular features for a hot-Jupiter planet have a contrast of 1 part to 10$^{3}$ while for Super-Earths planets around a M dwarf star the contrast is 1 part on \text10$^{4}$.
Hence, the precision on the inferred planetary radius, and thus on the stellar activity correction, needs to be of the same level.\\
Whilst stellar activity can reveal itself in multiple ways, a simple approximation is a combination of two different timescales: a long and a short one.
Given that we are analysing a quarter at a time, the purely technical upper limit of the long term is $\sim$90 days while the lower limit, for both long- and short-term, corresponds to the integration time ($\sim$0.024 days). 
Depending on the causes of the stellar variation in each time-serie these timescales will weight differently, and can be estimated measuring the power spectrum of the signal. 
We report in Table \ref{targets} the averaged most powerful period computed for the available quarters of all the stars, excluding quarter 1 as it is shorter.
We used the statistical model described in \S \ref{sec:model}, to fit the lightcurves.
The combination of the short and long covariance-lengths allow us to have a best fit the small scale variations and to reduce the uncertainty over long gaps. 
Model consisting of two component kernel function constrains the probability distribution over the function space more accurately.
To demonstrate the importance of this detailed modelling of two component kernel function, we fitted the stellar trend of KIC 3835670 (Q-6) three times:
using a RBF covariance function with single component with a best long term trend, a single component with a best short term trend and the best two component kernel function (Fig. \ref{fig:covariancelengths}).
We find that the using a single RBF with long time scale causes underfitting: it is not possible to capture small scale variations.
However, in the regions where data is not present, the uncertainty on the reconstructed model is small.
Using a single short term trend causes the data to be fitted very well, but in the gap regions the uncertainty on the function is very large.
The two component kernel achieves both goals: models the trend very accurately on small scales as well as having small uncertainty on the function in the regions where the observations are missing.

To identify points belonging to events such as transits or flares, we identified points which were outliers from our model.
As mentioned in \ref{sec:model}, such events should fall outside of the model of stellar trend for at least few consecutive observations.
We identified  the outliers greater than 3$\sigma$ and isolated a number of data points adjacent to them by defining the maximum duration $t^{event}_{max}$ of the intensity fluctuations we are looking for (see \S\ref{sec:model}, point \ref{time_event}).
In order not to affect the amplitude of the sought event we temporally removed clumps of outliers from the lightcurves, and trained a new GP model only on the remaining data (hereafter \textit{inliers}). 
Figure \ref{raw_inliers} demonstrates the necessity of this step: it shows how the value the model is altered when the GP fit is applied to the all the data, including a transit event, compared to the fit using only on the \textit{inliers} (with transit removed). 

To demonstrate the performance of the full procedure, we apply it to the star KIC-2571238, quarter 6. 
Fig. \ref{Kep_inliers} shows the model (black dashed) created for this star by the GPSC technique using a the fit on the \textit{inliers} (red points).
Excluded outliers, which were not used for second GP training, are marked in blue.
Confidence intervals are marked in green lines. 
Bottom panel shows the residuals after subtracting the model from the data.
We obtained a de-trended lightcurve where the main stellar contribution is removed non-parametrically and the main astrophysical features are untouched (transits and flares, Fig. \ref{Kep_zoom},  \ref{flares} respectively).

Note that we can also apply the model evaluated on the Long Cadence (LC) data to the Short Cadence (SC) lightcurves. 
This give us high resolution star-detrended residuals (Fig. \ref{high_residuals}) and it is particularly useful for studies of short lived  events like flares and transits (Murphy 2012).
It would have been possible to run the GPs directly on the SC lightcurves (working on sub-quarter/monthly-download separately), but that is inefficient computationally (see \S\ref{sec:model}, point \ref{time}).
Also, the standard deviation of the noise in the LC lightcurves is smaller than in the SC ones, making it easier to capture the stellar-trend without suffering the effects of increased noise. 

\subsubsection{Extrapolation of the stellar flux}\label{sec:extrapolation}
The Kepler mission continually observed the same field of view for four years, yielding a large number of very neat and continuous lightcurves.
In the exoplanet characterisation field, however, this unique opportunity does not occur and the data are sparse in time.
Observations normally last from $\sim$3 hours before the transit to $\sim$3 hours after the transit. 
In some case part of planetary orbital phase have been constantly recorded (Knutson et al. 2007), but this is almost prohibitive in term of costs and time. 
For this reason we designed the GPSC to perform well in a more realistic situation of data sparse in time. 
Having established the good performance of GPSC on full Kepler lightcurves, we tested it in a situation of periodic monitoring.\\
To produce this periodic monitoring scenario we selected a subset of data from the Kepler lightcurves, so that they contain only data spaced out by different time intervals.
We kept the length of the observations fixed throughout this exercise: we chose a duration of 10 hours ($\sim$ 20 long-cadence data points) as it is a reasonable time to observe a planetary transit. 
It allows in fact a long out-of-transit baseline where to consider the stellar contribution. 
For each quarter we recreated a set of 7 lightcurves, first simulating 10 hours observations every day, then 10 hours every second day and so on
until we reached 10 consecutive hours observed every week. 
We want to test if GPSC is able to reconstruct the full stellar trend when only sets of 10 hours observation are available for training the method.
To do this, we create a GPSC model using only 10 hour periodic observations, $\mathcal{M}^{10}$, and compare it the GPSC model created from the full data (using inliers only, $\mathcal{M}^{full}$).\\
We run the GPSC, as described in \S \ref{sec:detrending}, on each of the subsets of data, and extrapolate the $\mathcal{M}^{10}$ to all points in $\mathcal{M}^{full}$.
In this case we did not remove the data edges around the gaps and, most importantly, instead of only fitting the available data we extrapolated the model all over the time series, gaps included.
We then computed the standard deviation of the residuals of these two models.
The obtained value corresponds to the error on our extrapolation, assuming that the model extrapolated on the full time-series is the \textit{true flux} of the star. 
For each target, we processed all the quarters and we averaged the retrieved standard deviation value for the sparse datasets that have the same time interval.
Results are reported in Table \ref{tab:accuracy}.

\section{Results and discussion}
\label{results}
 
We analysed here the long cadence (LC) data of multiple Kepler quarters for a sample of stars with different stellar activities and effective temperatures $T_{eff}$ (Table \ref{targets}). 
We have been able to model simultaneously the stellar luminosity along with the instrumental systematics, disentangling the astrophysical features from the main star modulations. 
Such astrophysical features, like transits and flares, are recognisable in the lightcurve as sudden variations in the intensity.
Figures \ref{Kep_inliers}, \ref{Kep_zoom} and \ref{flares} show residuals after removing the stellar modulation and instrumental signatures.
There is no visible correlation or structure in the residuals, indicating that these trends were removed successfully.
Furthermore, if approximately correct noise standard deviation is supplied to the algorithm, GPSC will not overfit the noise, as the predictive marginal likelihood method will penalise overfitting sets of kernel hyperparameters (see GPML documentation for details about calculation of this likelihood).
The procedure of outlier identification seems to capture only the desirable clumps of outliers due to events, which we verified by visual inspection of all the fitted models.
Next, we resampled the same model to the short cadence (SC) data. 
The resolved LC and SC cadence residuals can then be used to confirm multiplanet systems via precise transit timing variation measurements (e.g. Ford et al 2012, Steffen et al 2013).

In a scenario of periodic observations that span different time intervals, we applied the GPSC to predict the stellar flux where data are not available. 
In the case of 10 hours of observations per day the extrapolated model is average accurate at the 10$^{-5}$ level compared to the one predicted with all the data.
When observations are sparse e.g. 10 hours every second, third, fourth and fifth day the measured accuracy is $\sim10^{-4}$, while every sixth and seventh it drops to 10$^{-3}$.
Table \ref{tab:accuracy} shows the accuracy of our stellar fit for each target. 
We notice that the accuracy of the fit also depends by the original noise standard deviation of the data.

\noindent We expect the ability of extrapolating the true curve to be a function of the ratio \textit{R} between the regularity of the observations and the stellar rotation period. 
When $\textit{R} \textless 1$ the GPSC is able to reconstruct the curve, however the more the ratio decreases, the better accuracy is achieved.
Accordingly, when $\textit{R} \ge 1$ we can expect a loss of information. 
This means that, to reach the same accuracy in the extrapolation, the time span between observations can be larger when observing a star with a slow rotation than when observing a star with a faster rotation.

\subsection{\textit{Technique validation}}
 \label{validation}
 \subsubsection{Test on real data}
 To test the GPSC technique we modelled the long cadence data, quarters Q 3-6, of Kepler-19 (Ballard et al. 2011). 
 We then resampled the same model retrieved for each different quarter Q, on the corresponding SC quarter. 
 In this way we obtained LC and SC residuals where the planetary transit of Kepler-19b is isolated. 
 Next we folded the residual time-series over the transit feature and we separately fitted each transit lightcurve with a Mandel $\&$ Agol model  (Mandel $\&$ Agol 2002) using a Markov chain Monte Carlo (MCMC) technique (Haario et al. 2006). 
 The stellar system and the limb darkening parameters implemented in the fitting function are taken from the discovery paper of Kepler-19b (Ballard et al. 2011).
 Figure \ref{LC_SC} shows the weighted mean depth with respective uncertainties computed for both LC and SC data for each quarter. 
 All the LC and SC measurements are consistent within themselves  at 1$\sigma$ level. 
 As expected the uncertainties of the LC depth are larger than the SC ones. 
 Furthermore we calculated the weighted mean of all the values of the SC depth obtaining ~
 $\overline{d}$ = 5.62$\cdot10^{-4} \pm 4.29\cdot10^{-6}$, a value which is consistent with Ballard et al. (2011) at 1-$\sigma$. \\

 \subsubsection{Test on simulated data}
We randomly chose KIC-1025967 (R$_{S}$ = 0.788 R$_{\odot}$)\footnote{http://exoplanetarchive.ipac.caltech.edu} from the Kepler catalogue as a star with no planets. 
We simulated a hot-Jupiter transit (R$_{P}$ = 1.03 R$_{J}$) with period P$\sim$7 days and we injected it along all the 16th available lightcurves of the star. 
We then proceeded with the GPSC and computed the final residuals. 
As previously done with KIC-2571238 we folded the residuals and fit each transit with the Mandel $\&$ Agol model using the MCMC technique. 
The measured planetary radius (R$_{hj}$ = 1.03 $\pm$ 5.19 $\cdot 10^{-3}$ R$_{J}$) differs from the original by 6.57 $\cdot 10^{-4}$ R$_{J}$. 
The same procedure was carried with a super-Earth simulated transit (R$_{P}$ = 0.26 R$_{J}$, P $\sim$ 15 days) where the retrieved planetary radius (R$_{se}$ = 0.26 $\pm ~$1.5 $\cdot 10^{-3}$ R$_{J}$) deviates from the original by 4.47 $\cdot 10^{-4}$ R$_{J}$.  
Hence the difference between the depth values amount at 2.20 $\cdot 10^{-5}$ and 3.78 $\cdot 10^{-6}$ for the hot-Jupiter and super-Earth transits respectively, proving the accuracy of the GPSC.

\section{Conclusions}

We introduced \textit{Gaussian Process method for Star Characterisation} (GPSC), a new non-parametric technique for modelling stellar flux variability and instrument systematics.
We tested the technique to the very stable and continuous Simple Aperture Photometry data of Kepler, in order to investigate the stellar-activity in the visible band, but it can be applied to similar datasets.
The GPSC allows to disentangle temporal events, such as flares, planetary transits and sudden intensity variations, from the long-term stellar modulation.
Additionally, it is able to reconstruct the stellar brightness variations in the scenario where observations are taken with time intervals.

After modelling the stellar variability, the detrended residuals are sampled both in long cadence and short cadence mode, i.e. with an integration time of 29.4 min and 58.9 s respectively. 
We tested the GPSC on a known star, KIC-2571238 (Kepler-19), by fitting a transit model lightcurve to the residuals with a Markov Chain Monte Carlo approach. 
The measured value of the depth is consistent with the one published in the discovery paper (Ballard et al 2011). 
Additionally we injected a hot-Jupiter transits in all the available lightcurves of KIC-1025967, a star without planets. 
As done for KIC-2571238 we proceeded with the GPSC and we fit the residuals to measure the transit depth. 
We repeated the same for a transiting super-Earth scenario.
The retrieved depths are consistent with the values used for the simulation at the $10^{-5}$ level, further validating the functionality and accuracy of our technique. 
The GPSC was able to model all the instrumental systematics and the stellar signal in a non-parametric way, without distorting the signal or injecting noise as it may happen with a simple least square fit; moreover it derived  an absolute calibration within the events that reside in the same quarter.

The GPSC can also be used to extrapolate the long term modulation of the stellar flux in a situation of non-continuous monitoring of the target.
For each target analysed we recreated a scenario where only 10 hours observations, spaced out by different time intervals, are provided. 
We also showed the possibility to extrapolate the stellar flux correctly at the $10^{-4}$ level.  
This accuracy is pivotal for the study of the atmosphere of exoplanets or for detecting very small planets.

\centering 
\begin{table}[bc]

 \centering
 \begin{tabular}{ | c | c | c | c | c | c |}
 \hline \hline
 \multicolumn{6}{|c|}{Targets} \\
 \hline \hline
 KIC & T$_{eff}$ [K]  &  R [R$_{\odot}$]  &  Kep$_{mag}$ & Flag &$\Delta$P [$\%$] \\
 \hline 
  3835670 &  5642  &  1.790 &  13.397  &Planetary candidate & 80.4 $\pm$ 20.3\\  
  2571238 &  5541  &  0.850  & 11.898 &  Exoplanet (Kepler-19b) & 88.8	$\pm$ 11.7\\ 
  6291653 & 5236 &	0.095 &15.305 & Planetary candidate & 89.6	$\pm$16\\ 
  1026957 & 4917 & 0.788 & 12.559 & False positive & --\\
  7700622 &  4789  &  0.649   & 12.968 & Planetary candidate & 69.6 	$\pm$32.4\\  
  10748390 &	4766	 &	0.058 & 9.147 & Planetary candidate & 28.5$\pm$	8.8\\
  3128793 &  4668 &  0.637 & 14.633 &  Planetary candidate & 70.9	$\pm$  9.5\\  
  7603200 &  3900  &  0.612 & 12.925 & Planetary candidate & 81.4	$\pm$	27.8\\  
  7907423 &  3803  &  0.492  &  15.234 & Planetary candidate & 89. $\pm$	14.8\\  
  5080636 & 3673 & 0.024 & 14.404 & Planetary candidate & 38.3$\pm$	18.4\\
 \hline
 \hline
 \end{tabular}
 \caption{Table of the targets. For each target we report here the effective temperature ($T_{eff}$), the stellar radius (R$_{\odot}$), the Kepler magnitude (K$_{mag}$), 
 the flag as in the KOI catalogue and the variation on the most relevant period of the available quarters ($\Delta$P).}
 \label{targets}
\end{table}

\centering 
\begin{table}

 \centering
 \begin{tabular}{ | c | c | c | c | c | }
 \hline \hline
 \multicolumn{5}{|c|}{Accuracy of the extrapolated model [$\%$]} \\
 \hline \hline
  & 3835670  & 6291653  & 7700622 & 10748390  \\
  \hline
  10hr/1d & 0.032 & 0.041 & 0.034 & 0.069 \\
  10hr/2d & 0.037 & 0.046 & 0.036 & 0.075  \\
  10hr/3d & 0.043 & 0.054 & 0.043 & 0.081  \\
  10hr/4d & 0.046 & 0.058 & 0.044 & 0.084  \\
  10hr/5d & 0.049 & 0.064 & 0.049 & 0.125  \\
  10hr/6d & 0.055 & 0.069 & 0.062 & 0.124  \\
  10hr/7d & 0.062 & 0.077 & 0.065 & 0.143  \\
  \hline
  \hline
  & 3128793 & 7603200 & 7907423 & 5080636  \\ 
   \hline
  10hr/1d & 0.092 & 0.036 & 0.036 & 0.169\\
  10hr/2d & 0.132 & 0.045& 0.045 & 0.056 \\
  10hr/3d & 0.153 & 0.058  & 0.065 & 0.069\\
  10hr/4d & 0.158 & 0.064 & 0.068 & 0.069\\
  10hr/5d & 0.197 & 0.080 & 0.084 & 0.085\\
  10hr/6d & 0.302 & 0.103 & 0.102 & 0.102\\
  10hr/7d & 0.321 & 0.117 & 0.118 & 0.143\\
  \hline
  \hline
 \end{tabular}
 \caption{Table of the average accuracy of the GPSC extrapolated model compared to the model fit on the complete lightcurve. 
 The average has been measured over all the available long-cadence quarters, excluding quarter 1, in a scheme where we selected 10 hours observation per day, every second day, every third day, every fourth day, every fifth day, every sixth day and every week. 
 The targets are in decreasing order of temperature (Left to right, top to bottom). 
 The values in the table quantify the measurement $\vec{std} \left( \mathcal{M}^{10}(t_i) - \mathcal{M}^{full}(t_i) \right) $ where $\mathcal{M}(t_i)$ is the value of the luminosity prediction at time $t_i$ for a GPSC model, see \S\ref{sec:extrapolation}.
 }
 \label{tab:accuracy}
\end{table}

\begin{figure}
\centering
\label{fig:gpdemo}
\epsfig{file=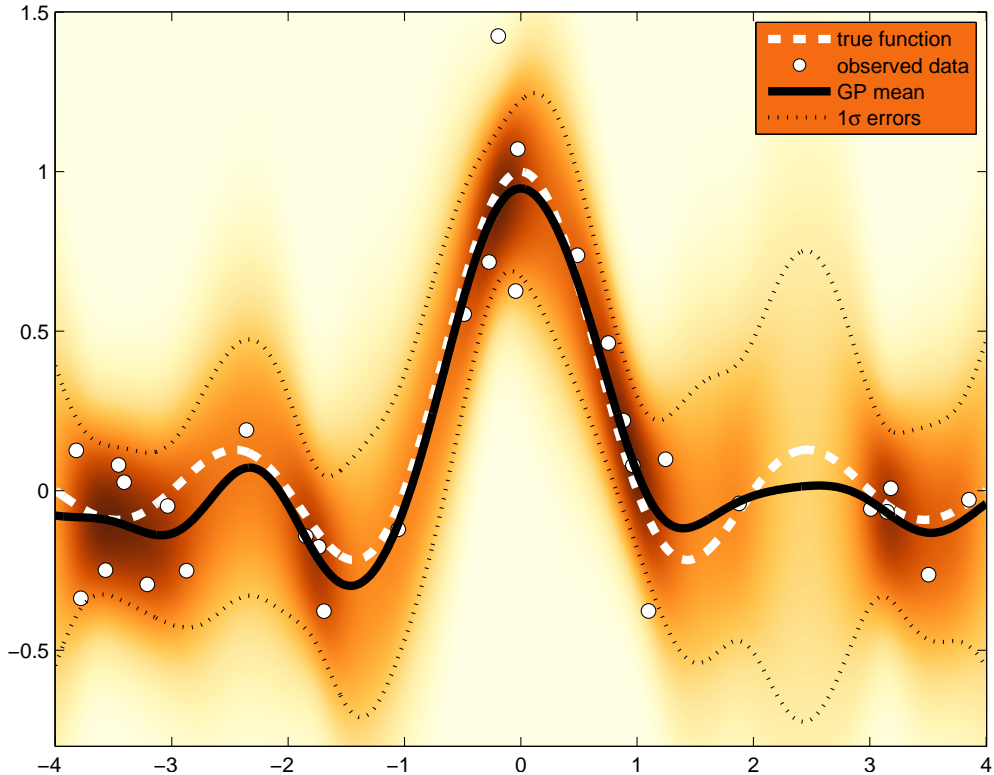,width=0.8\linewidth}
\caption{
Demonstration of Gaussian Process Regression. 
White points are noisy observations of underlying unknown function (white dashed), which in this example was a $sinc$ function. 
A posterior Gaussian Process creates a distribution on the function space: we condition it on the observed points and use a prior in a form of covariance function with RBF kernel and a zero mean function.
Black solid line is mean of the posterior distribution at any point, black dotted lines are 1-$\sigma$ errors and the intensity is a normalised probability distribution on the function space.
Note that in the region between values of $2<x<3$ there is no available observed points, which results in increased uncertainty on inferred function value in this region.
Opposite is true for region $-4<x<-3$: there are many observed points available which results in high confidence about the function value.
}
\end{figure}

\begin{figure}
\centering
\begin{tabular}{c c }
\epsfig{file=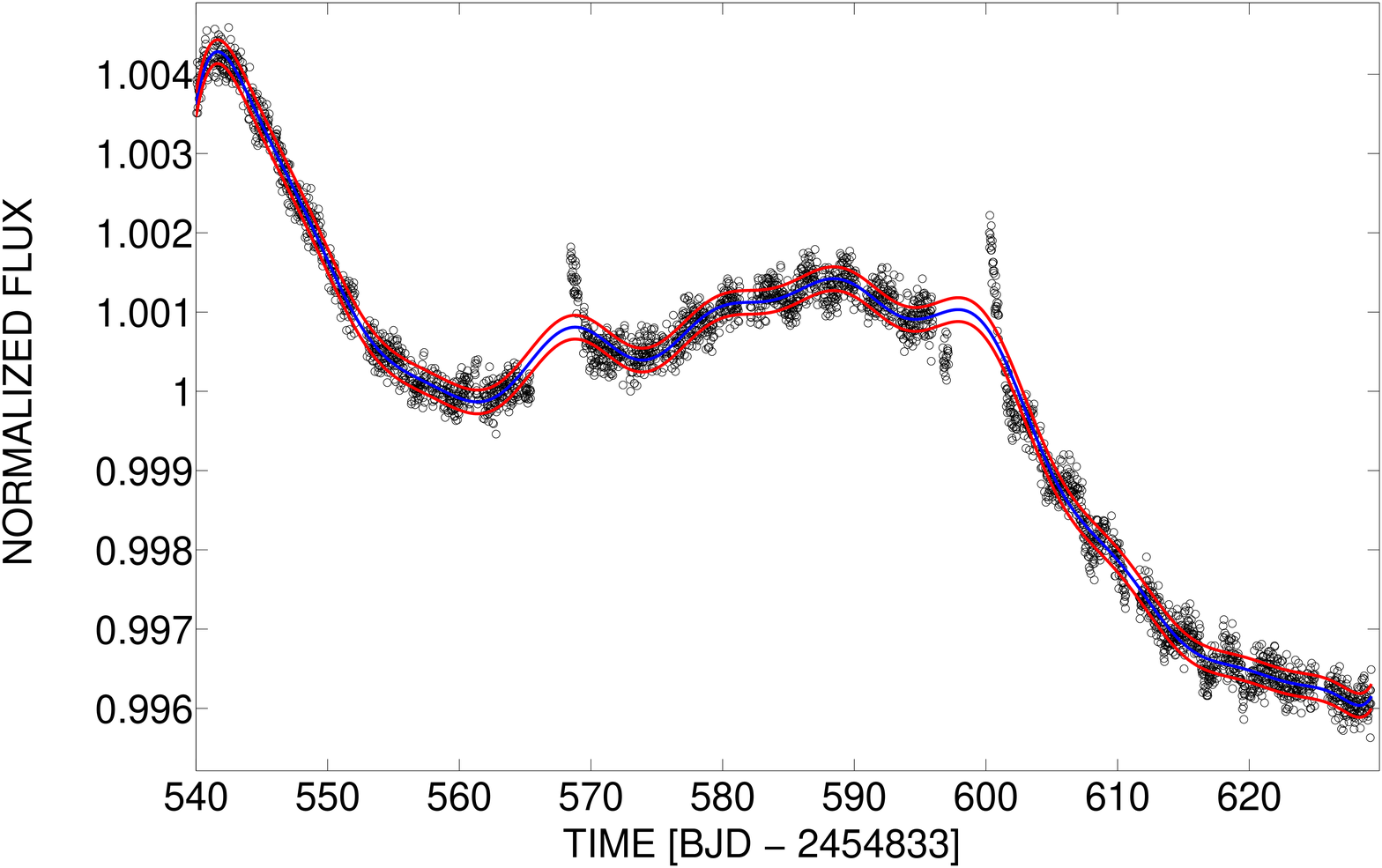,width=0.5\linewidth,clip=, trim = 5 10 80 40} 
\epsfig{file=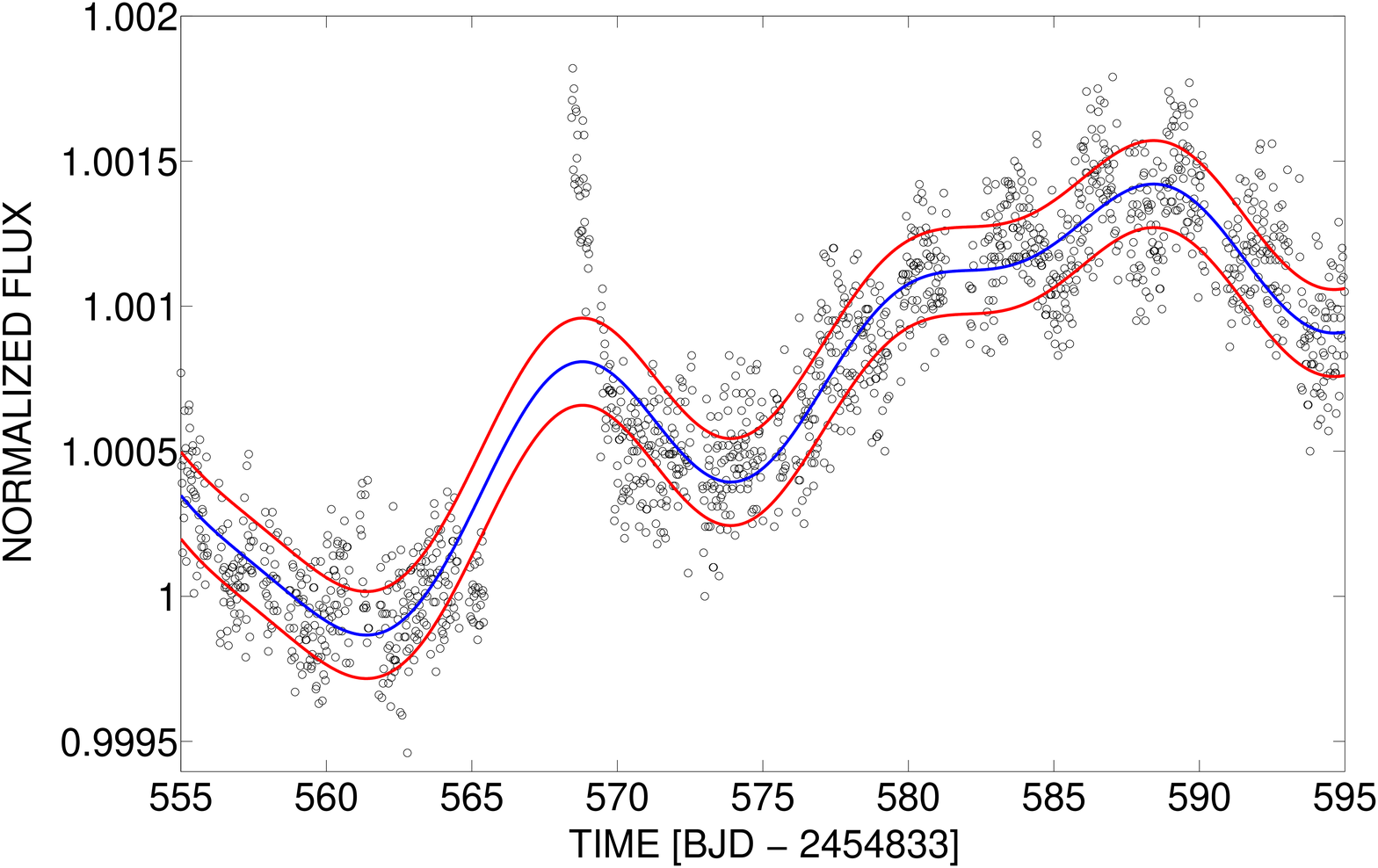,width=0.5\linewidth,clip=, trim = 5 10 80 40} \\
\epsfig{file=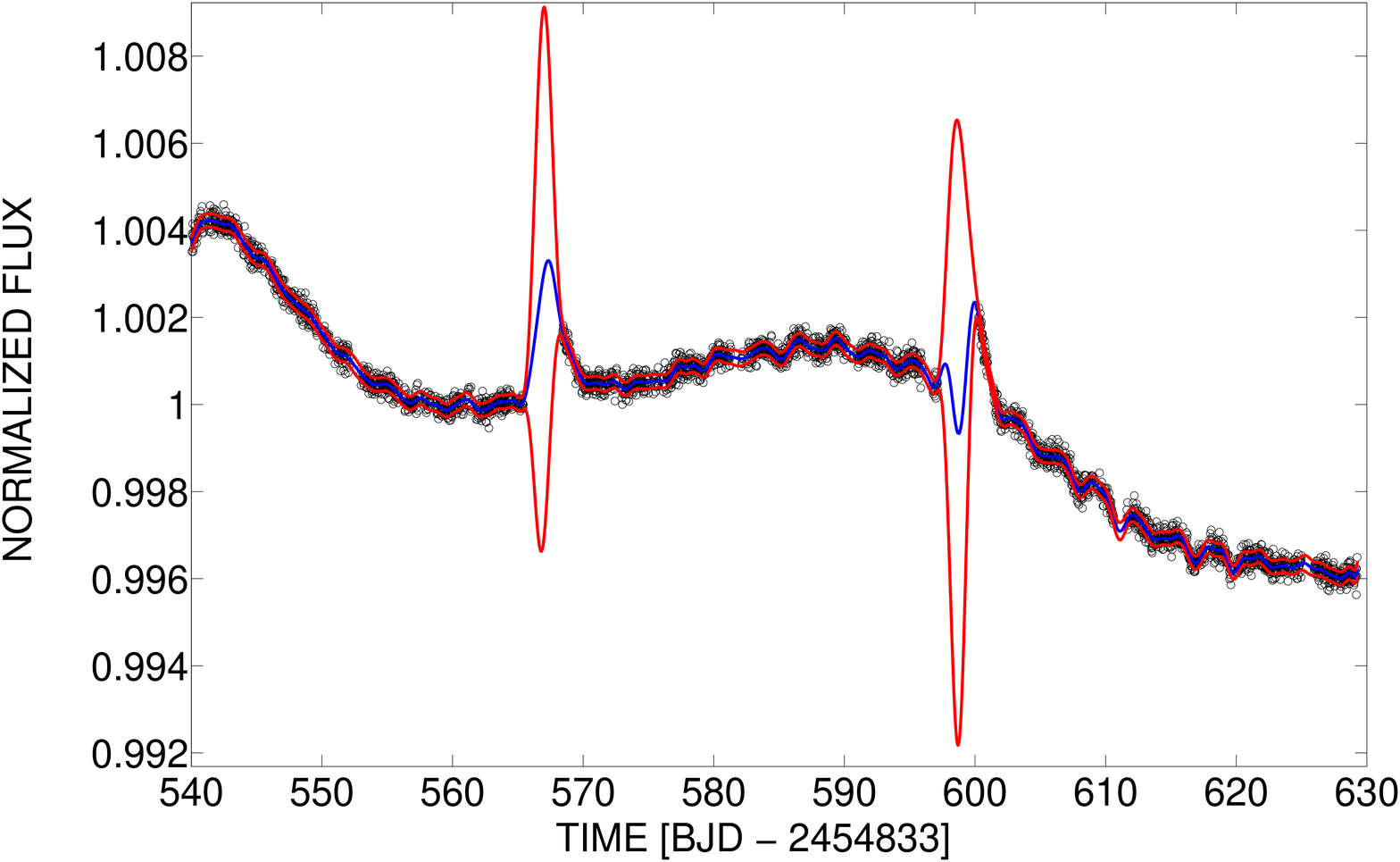,width=0.5\linewidth,clip=, trim = 5 10 80 40} 
\epsfig{file=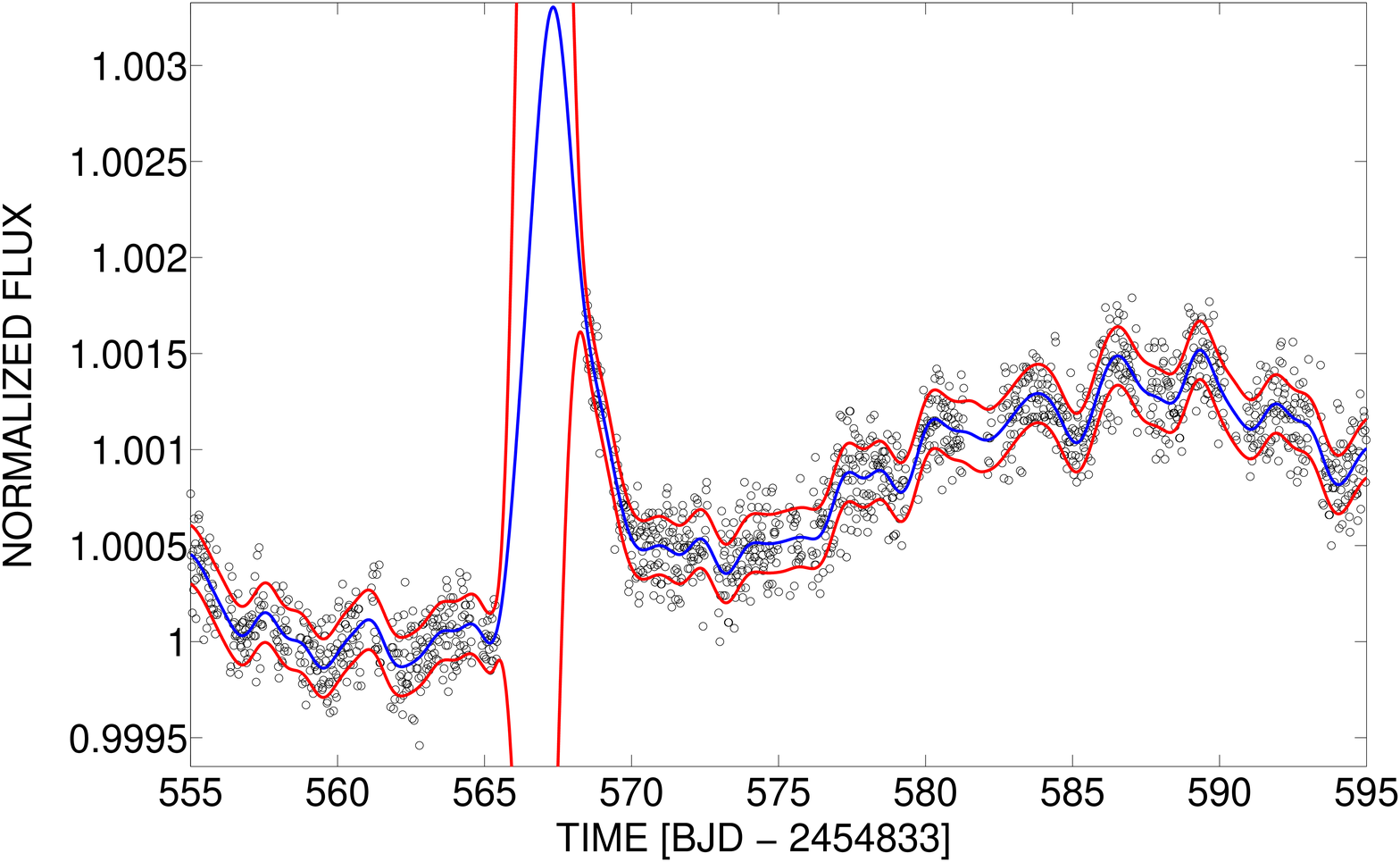,width=0.5\linewidth,clip=, trim = 5 10 80 40} \\
\epsfig{file=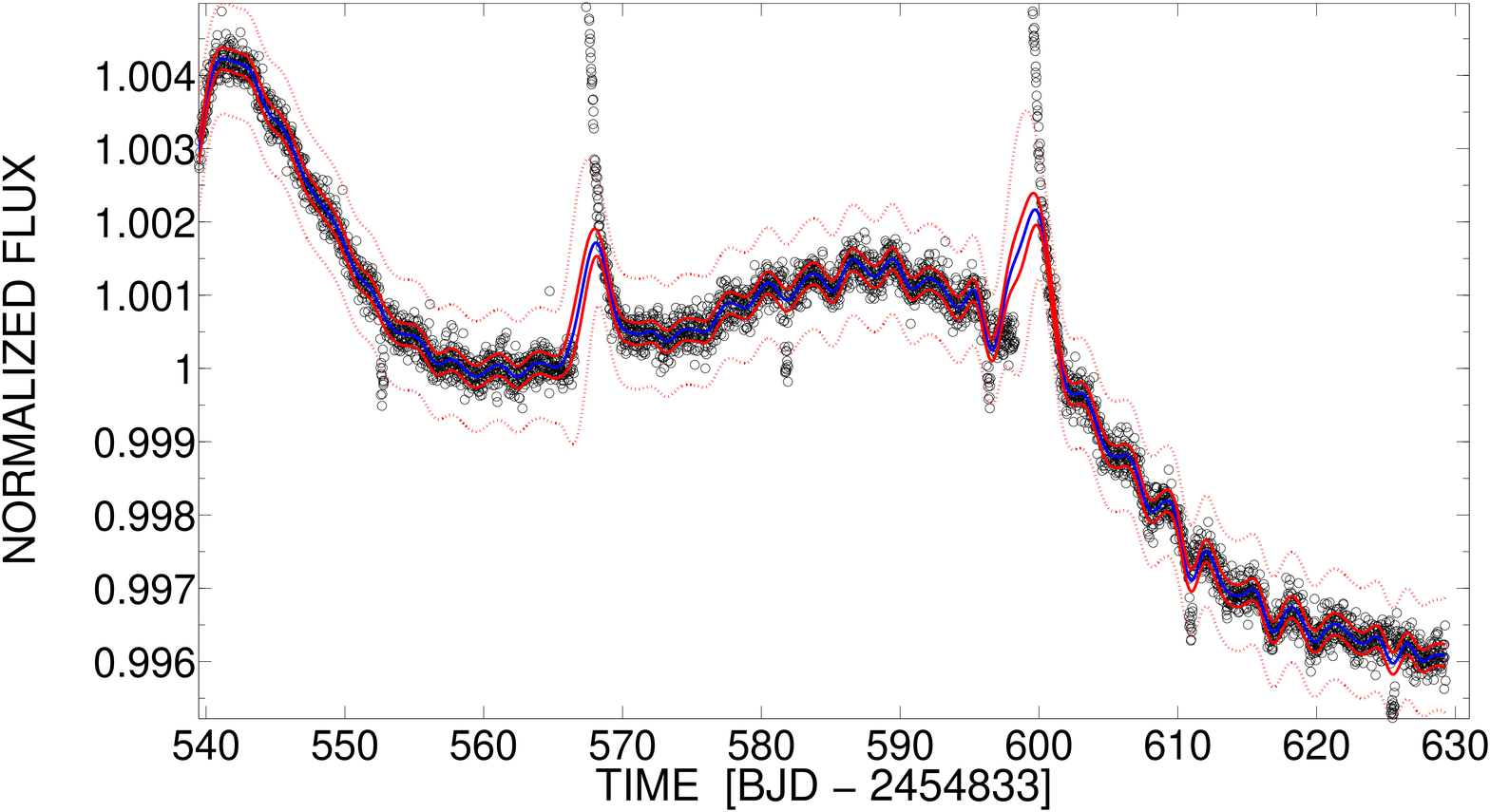,width=0.5\linewidth,clip=, trim = 5 10 80 40} 
\epsfig{file=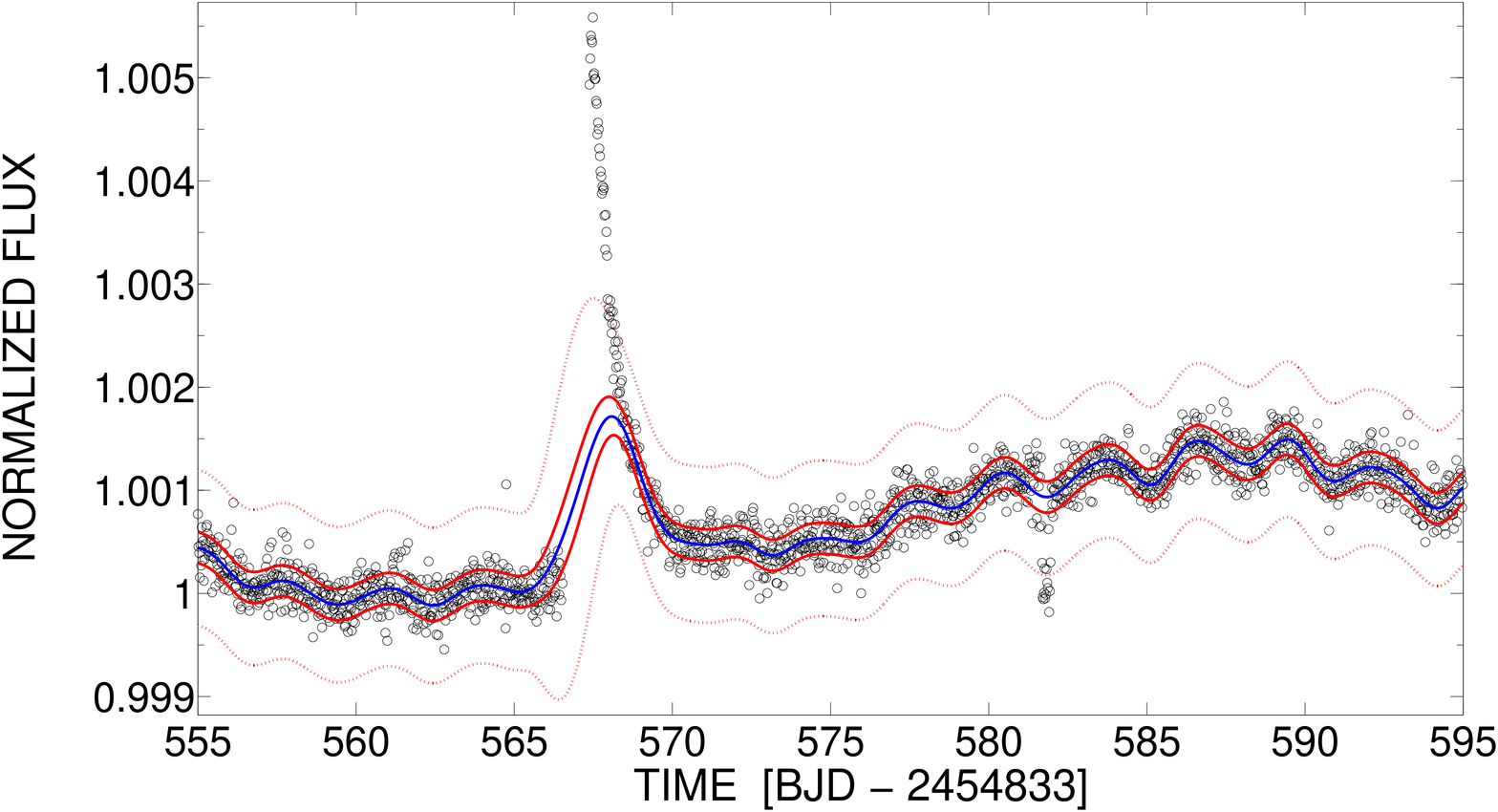,width=0.5\linewidth,clip=, trim = 5 10 80 40} \\
\end{tabular}
\caption{GPSC fit on the inliers of  KIC 3835670 (Q-6). \textit{Top:} Fit using only a long covariance-length (\textit{left}), and a zoom-in on the data (\textit{right}).
The long covariance-length gives better uncertainties over the gaps but misses the short-term fluctuations.. 
\textit{Middle:} Fit using only a short covariance-length (\textit{left}), and a zoom-in on the data (\textit{right}). The short covariance-length defines well the quick modulations but do not constrain well the uncertainties over the gaps. 
\textit{Bottom:} Fit using a combination of short and long covariance-lengths (\textit{left}) and zoom-in on the data (\textit{right}). The combination of kernels  describes well the short modulations
and properly constrains the uncertainties on the gaps.  The data are plot in black,  the blue line 
represents the mean model while the red line is the 1-$\sigma$ uncertainties.} 
\label{fig:covariancelengths}
\end{figure}

\begin{figure}
%\epsscale{0.7}
\plotone{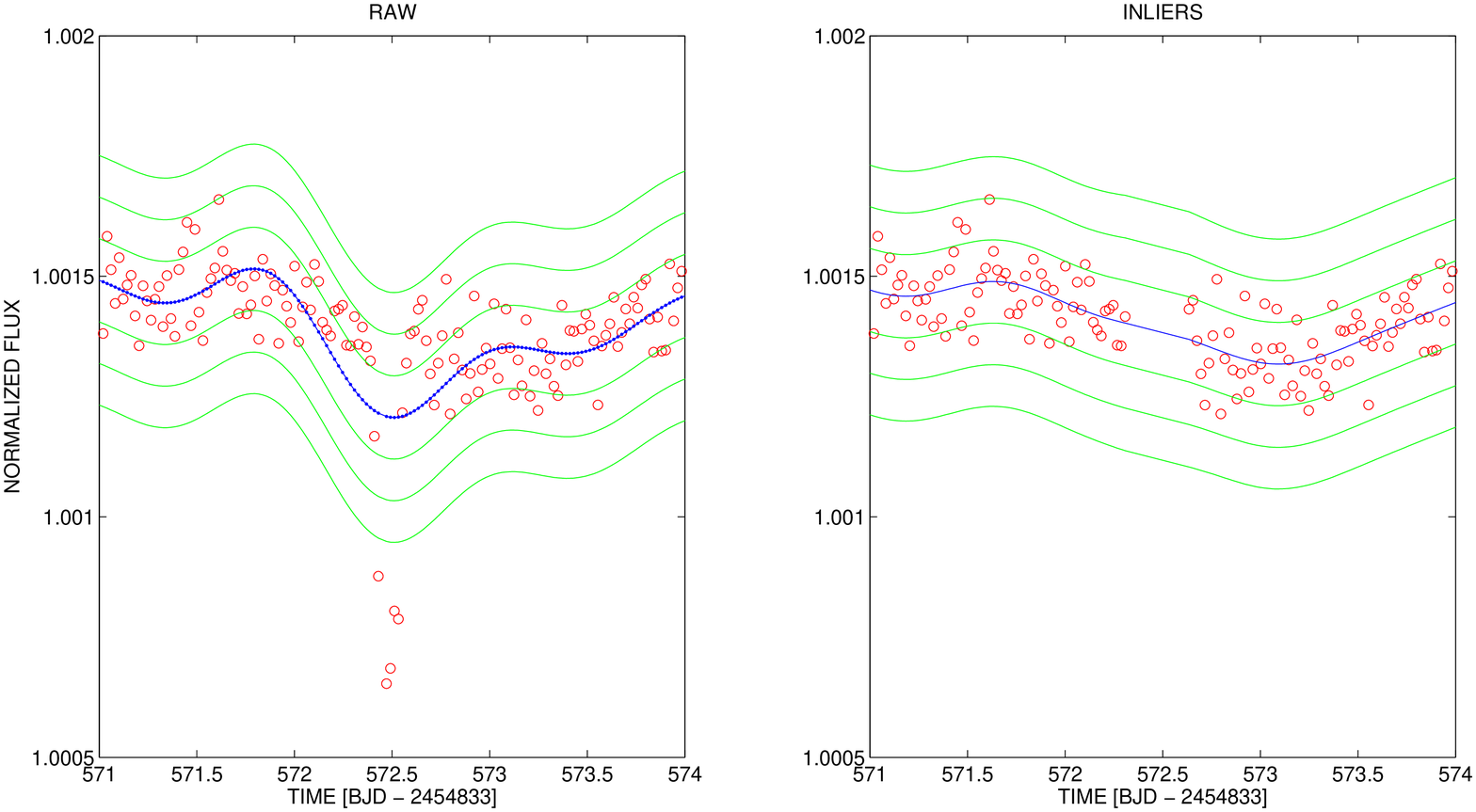}
\caption{GP model prediction on the totality of the data (\textit{raw - left}) and after the removal of the clusters of outliers (\textit{inliers - right}).  The data are plotted in red, while the model and
 the 1,2,3 $\sigma$ confidence limits are represented by the blue dotted line and green lines respectively.}
\label{raw_inliers}
\end{figure} 

\begin{figure}
%\epsscale{0.7}
\plotone{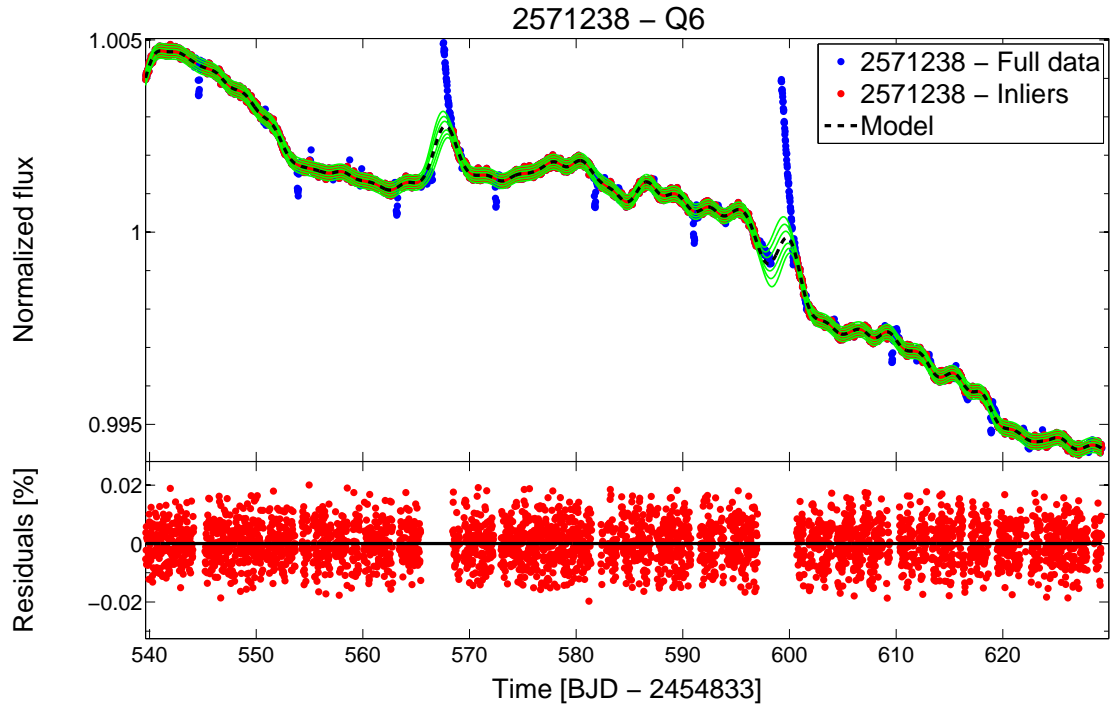}
\caption{GP model prediction (\textit{top}) and residuals (\textit{bottom}) on the long-cadence \textit{inliers} of Kepler-19 (Q-6). The inliers data are plotted in red, the outliers are plotted in blue,  the model and
 the 1,2,3 $\sigma^{*}$ confidence limits are represented by the black dotted line and green lines respectively.}
\label{Kep_inliers}
\end{figure}

\begin{figure}
%\epsscale{0.7}
\plotone{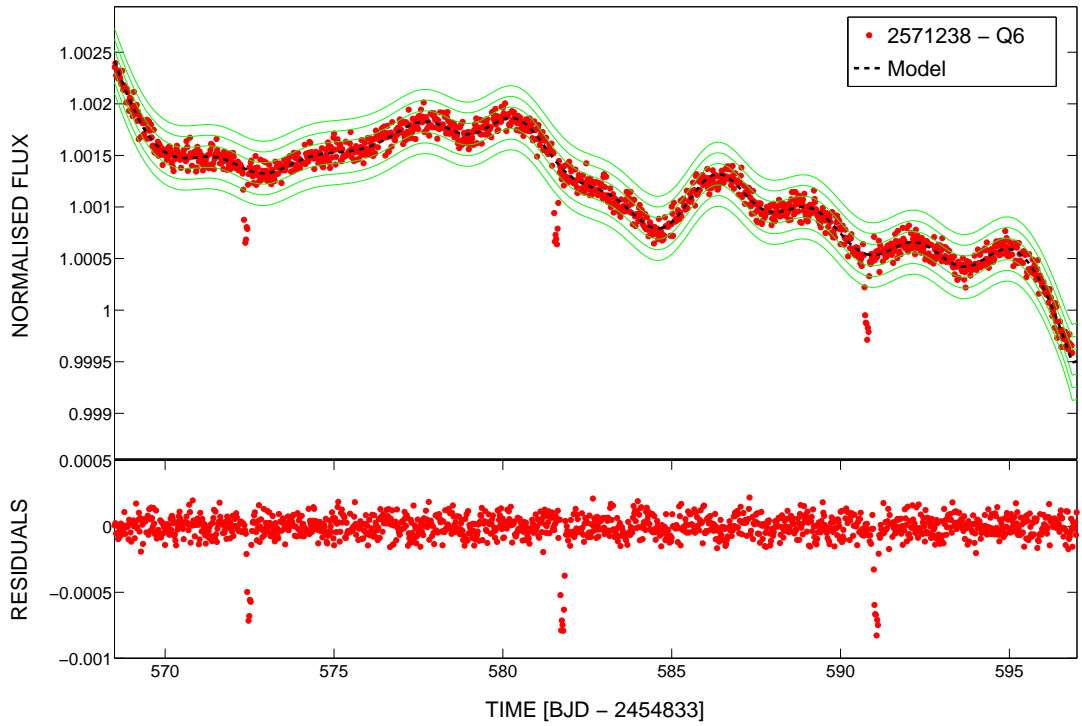}
\caption{GP model (black dotted line) applied to the 6th quarter long cadence data (red circles) of Kepler-19 and the respective residuals. The raw data are shifted for clarity.}
\label{Kep_zoom}
\end{figure} 

\begin{figure}
%\epsscale{0.7}
\plotone{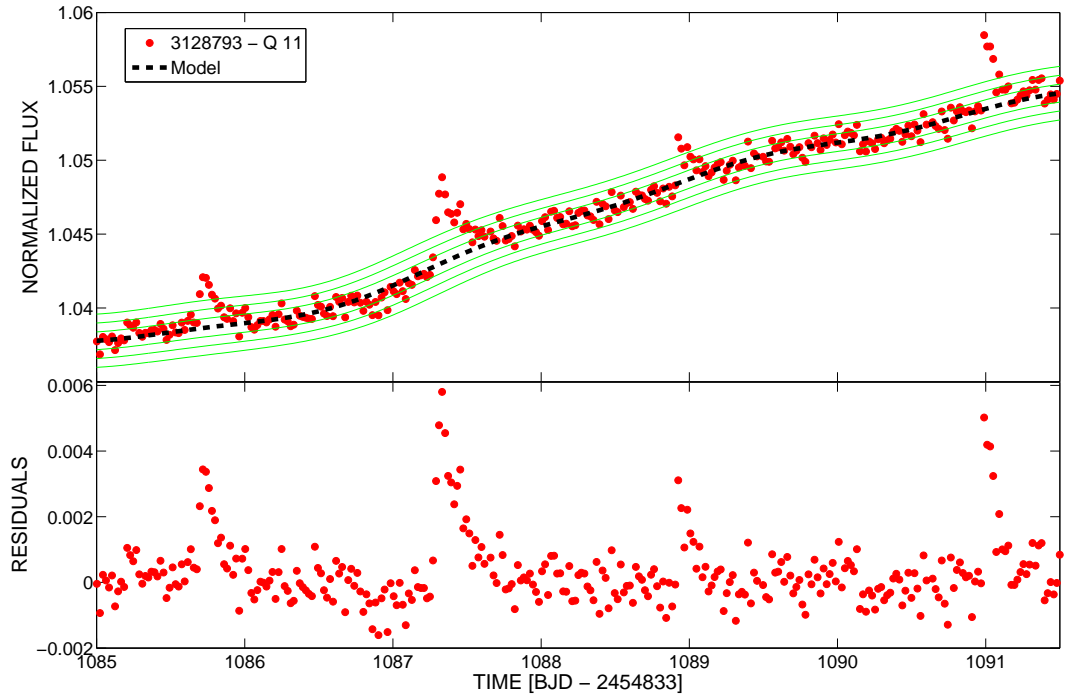}
\caption{Zoom-in on stellar flares of the long cadence observations of KIC 3128793 (Q11). In the top  panel we show the fit to the data while in the bottom panel we show the residuals after the GPSC detrend. The data are plotted in red, while the model and
 the 1,2,3 $\sigma^{*}$ confidence limits are represented by the black dotted line and green lines respectively.}
\label{flares}
\end{figure} 

\begin{figure}
\centering
\begin{tabular}{c}
\epsfig{file=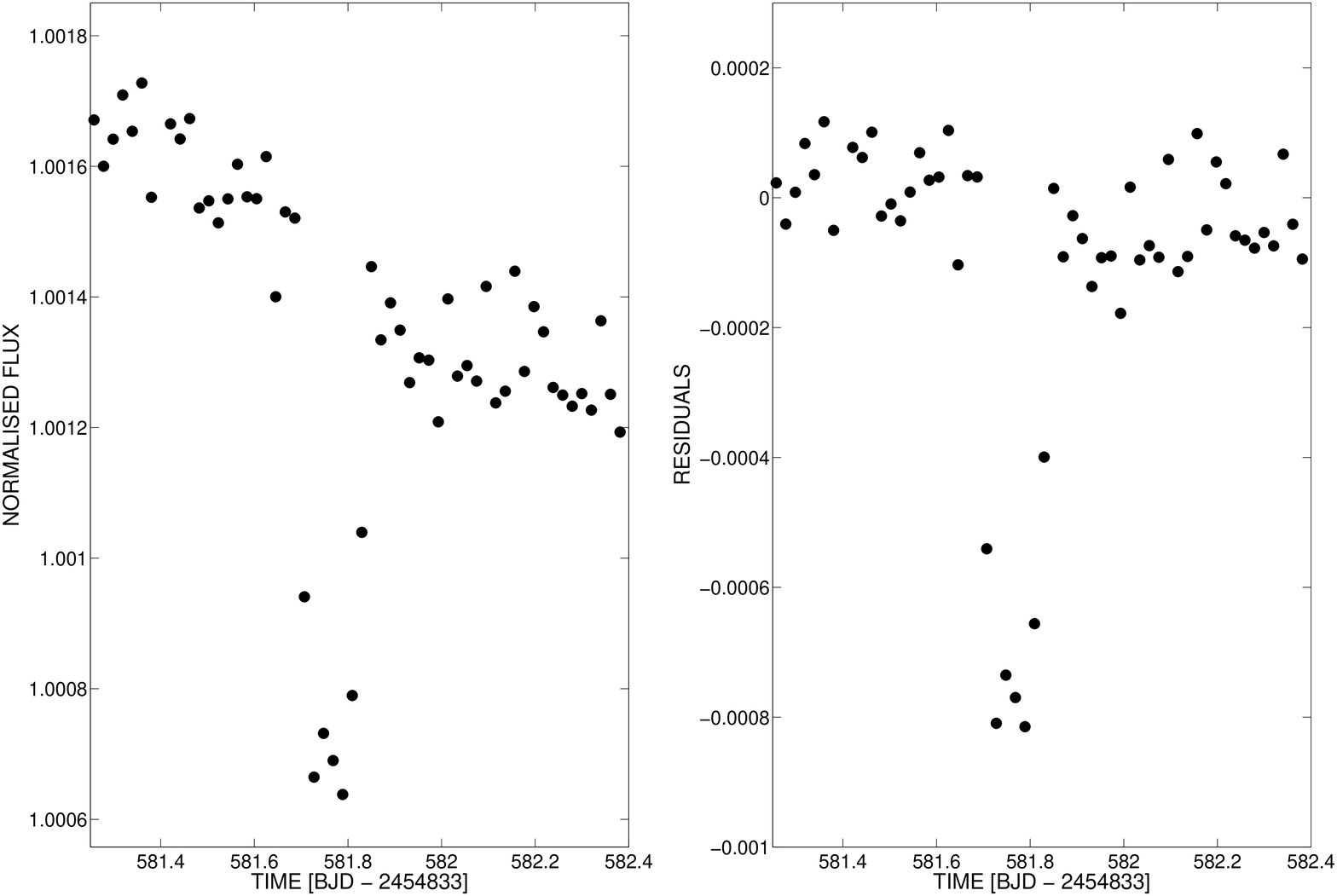,width=0.8\linewidth,clip=} \\
\epsfig{file=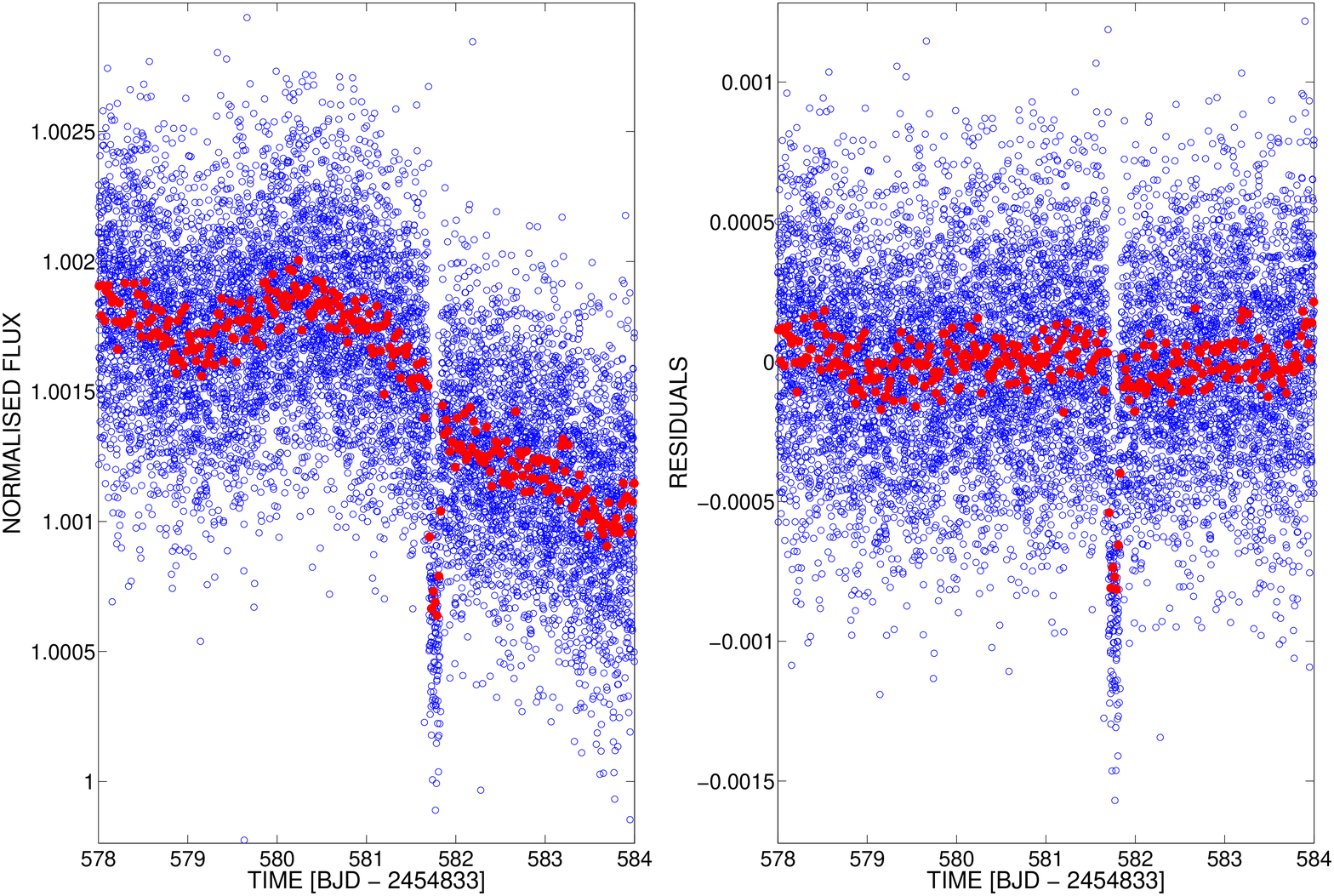,width=0.8\linewidth,clip=} \\
\end{tabular}
\caption{Top panel: Zoom-in of a transit of Kepler-19b (Q6) observed at long cadence before (\textit{left}) and after (\textit{right}) the GPSC detrend technique. Bottom panel: Zoom-in of the same transit of Kepler-19b (Q6) 
with the application of the GPSC model, retrieved for the long cadence observations (red dots), to the short cadence data (blue circles).} 
\label{high_residuals}
\end{figure}

\begin{figure}
\epsscale{0.9}
\plotone{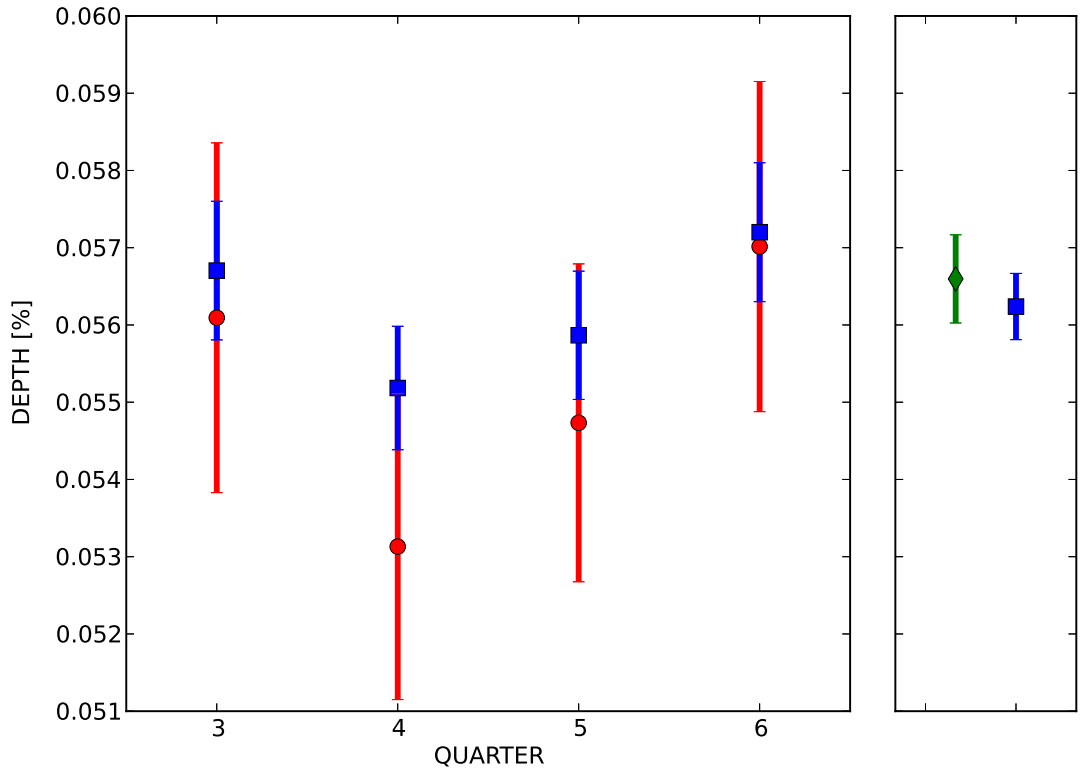}
\caption{\textit{Left:~} Mean transit depth values and uncertainties for quarters Q 3,4,5,6 computed for the long cadence data (red circles) and short cadence data (blue squares). \textit{Right:~} 
Mean transit depth of all the quarters (blue square) -- short cadence data -- compared with the transit depth by Ballard et al. 2011 (green diamond).}
\label{LC_SC}
\end{figure}

\begin{figure}
\centering
\begin{tabular}{c c}
\epsfig{file=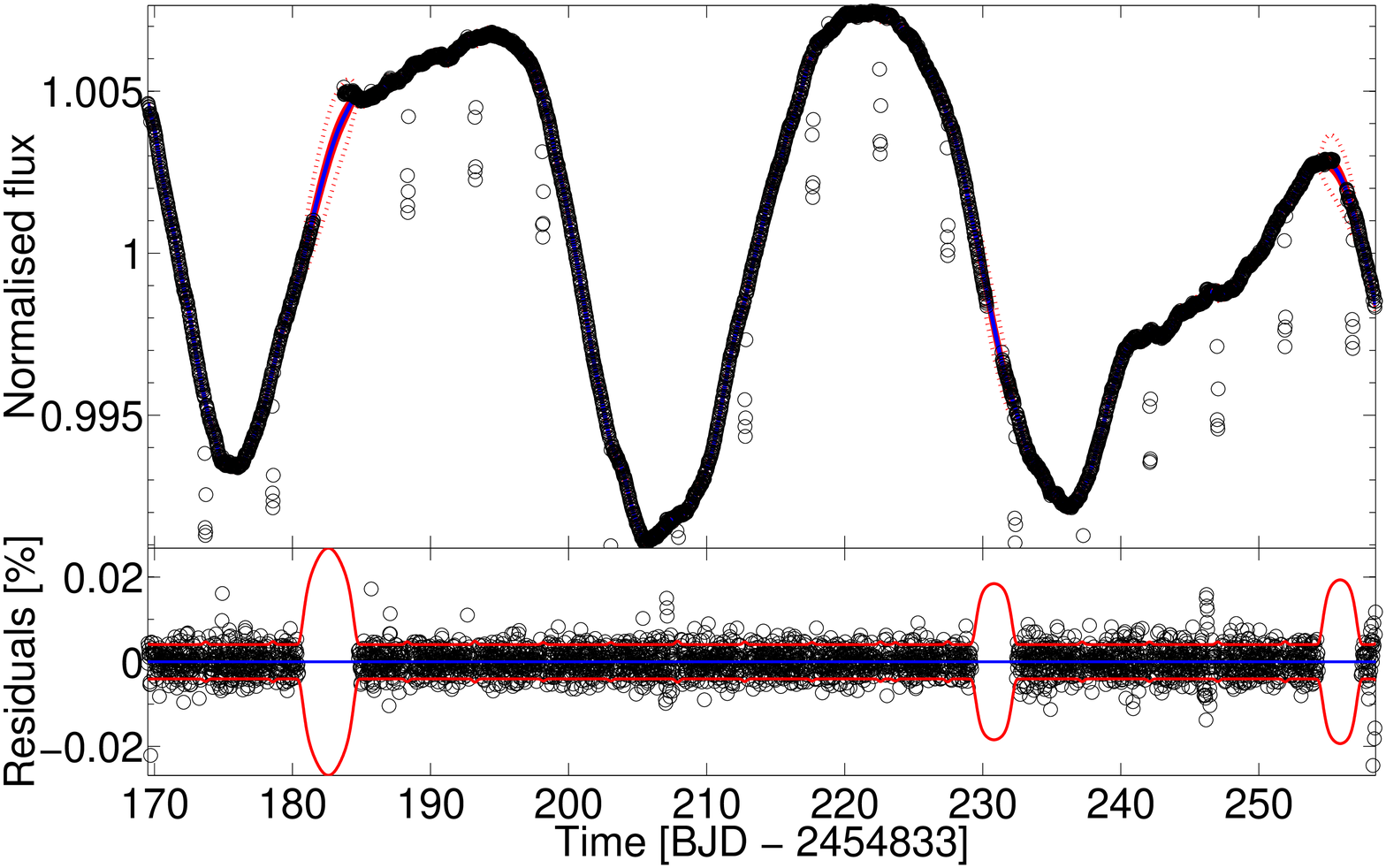,width=0.4\linewidth,clip=, trim = 10 0 80 20} 
\epsfig{file=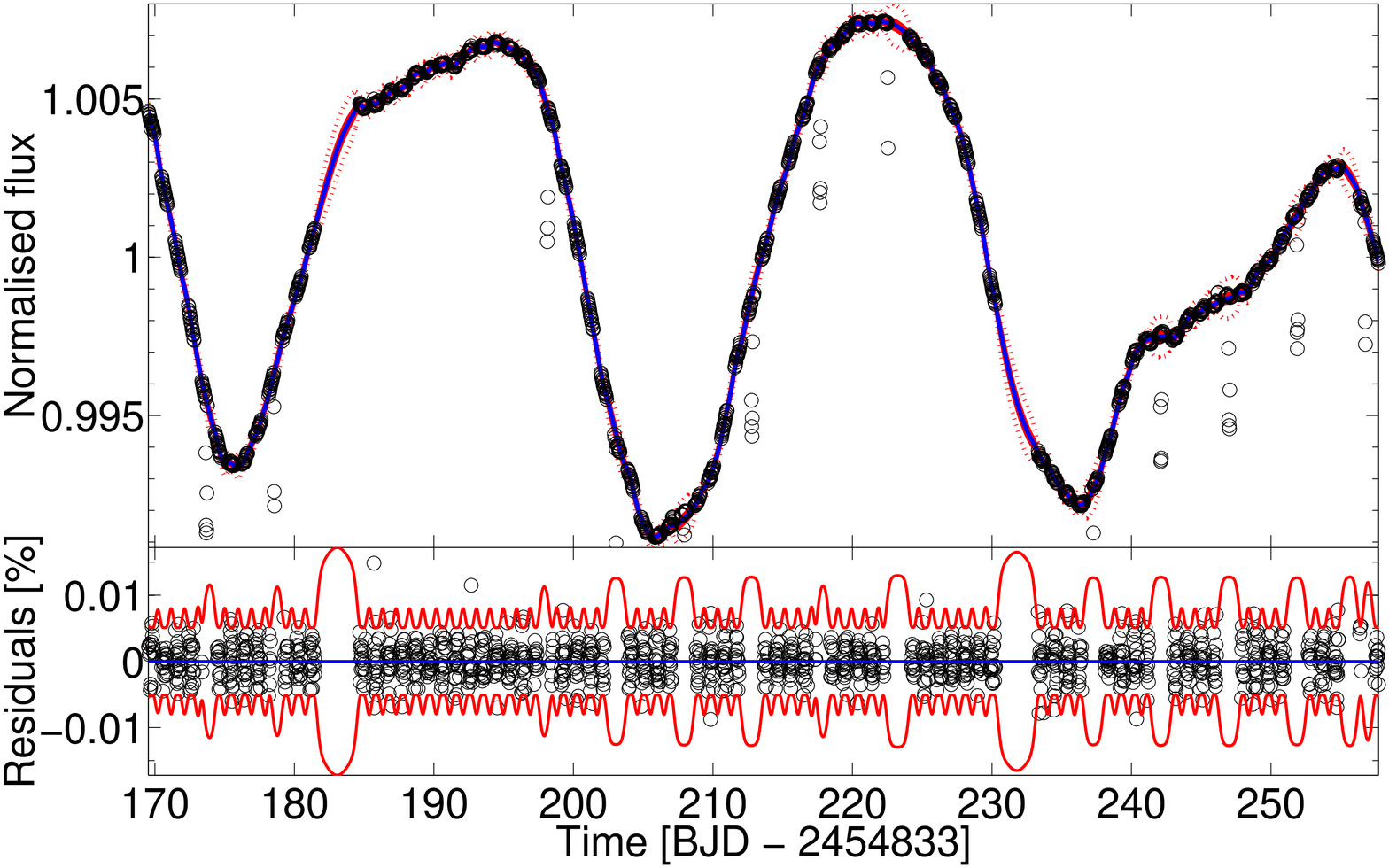,width=0.4\linewidth,clip=, trim = 10 0 80 20} \\
\epsfig{file=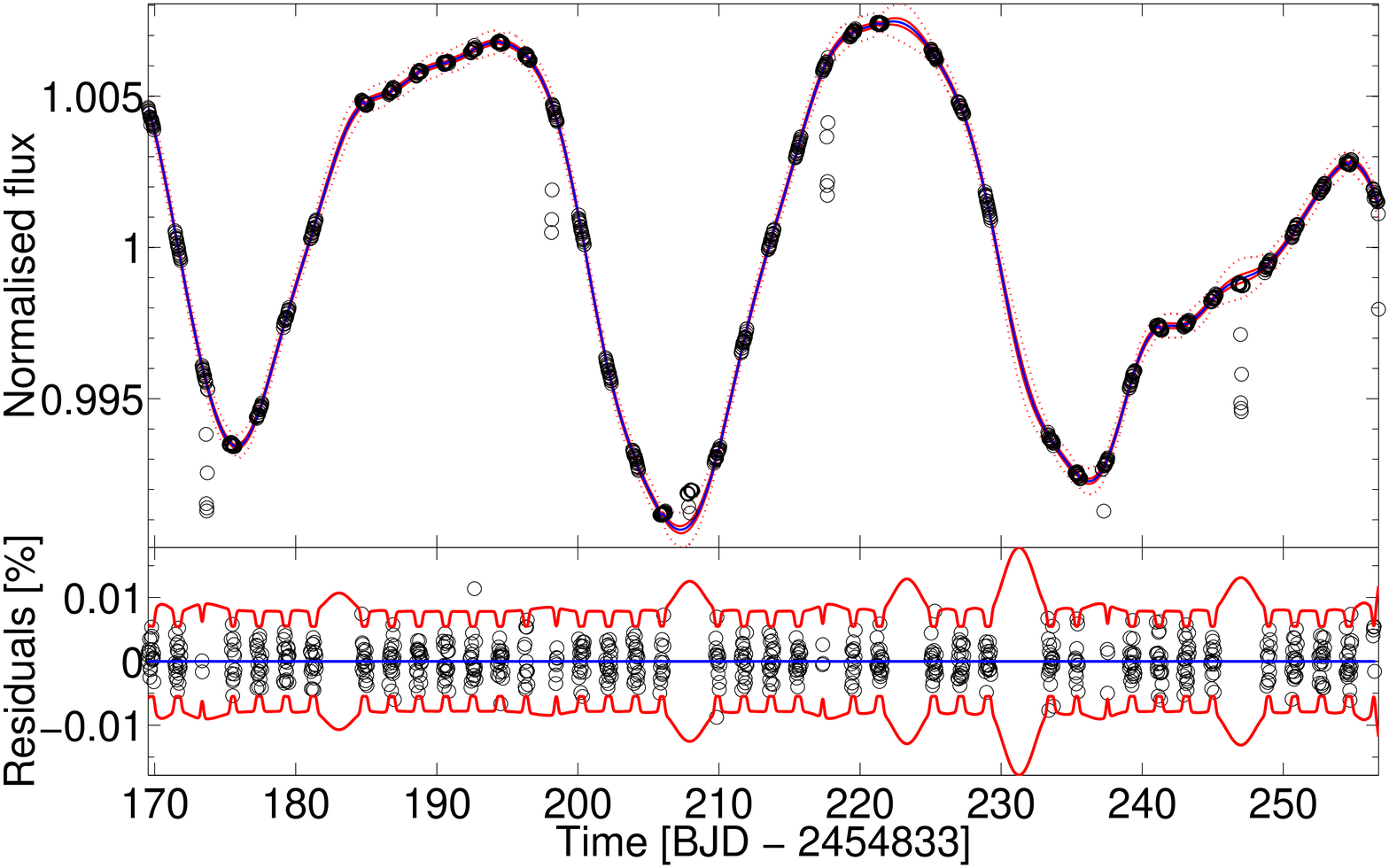,,width=0.4\linewidth,clip=, trim = 10 0 80 20} 
\epsfig{file=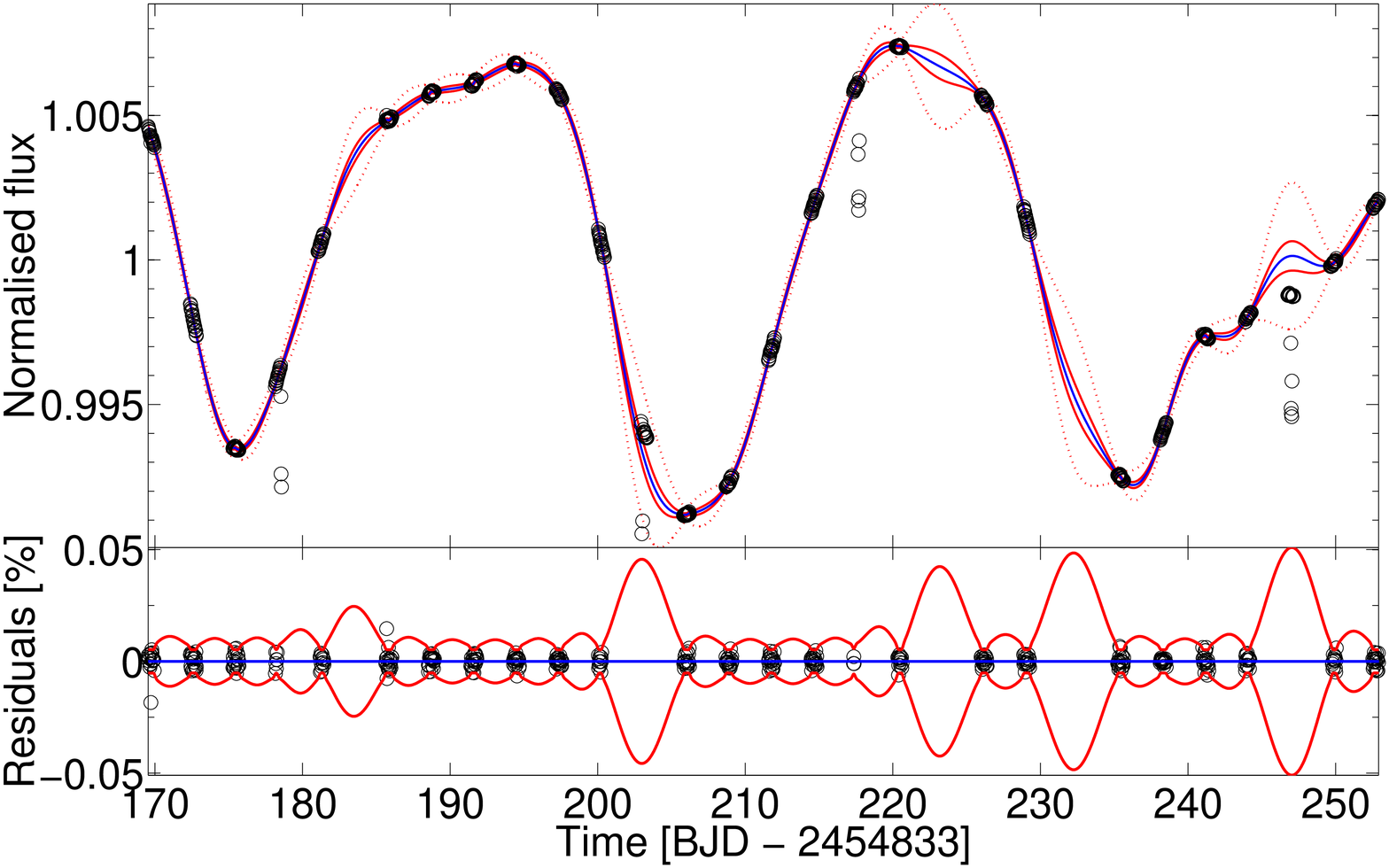,width=0.4\linewidth,clip=, trim = 10 0 80 20} \\
\epsfig{file=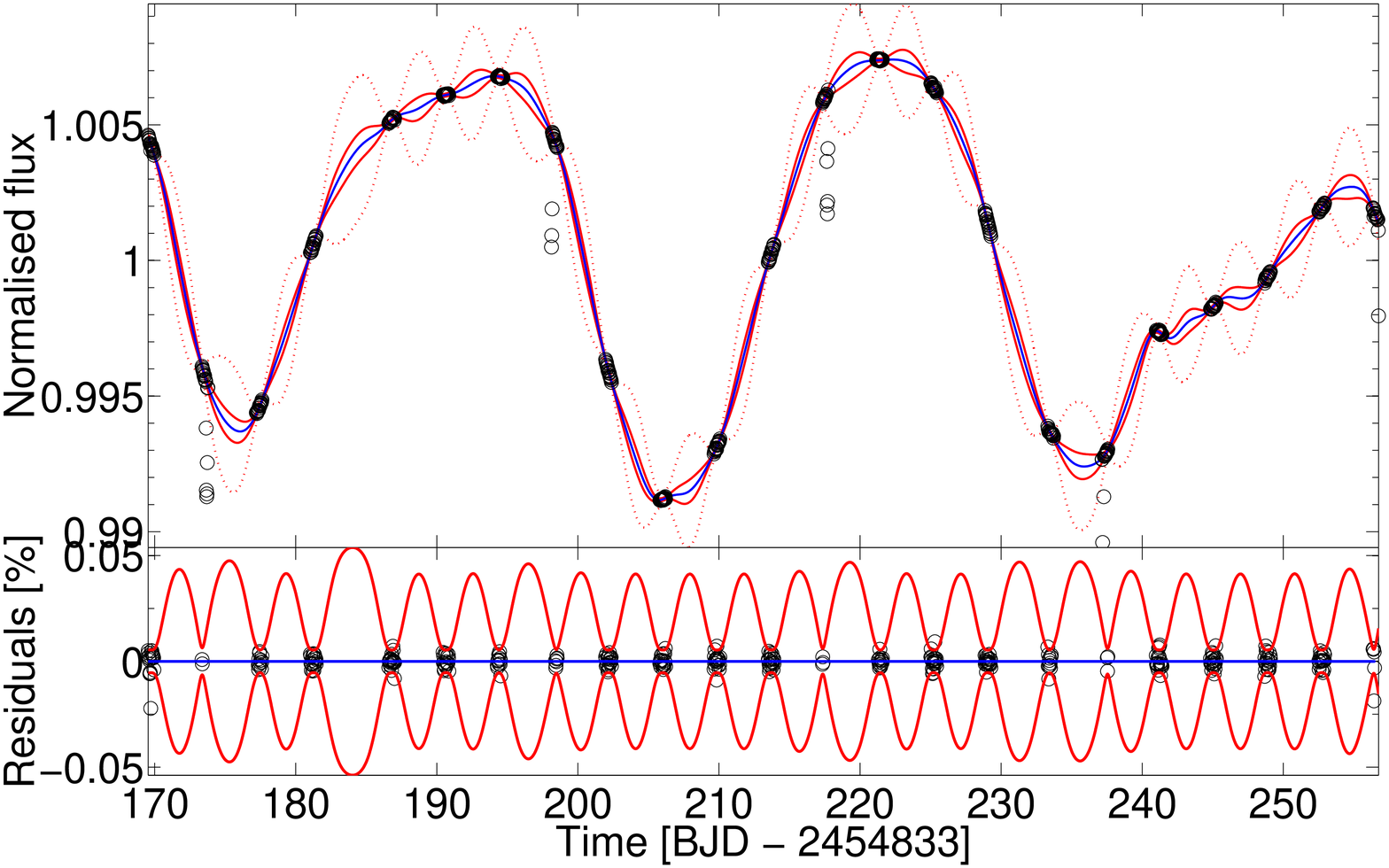,width=0.4\linewidth,clip=, trim = 10 0 80 20}
\epsfig{file=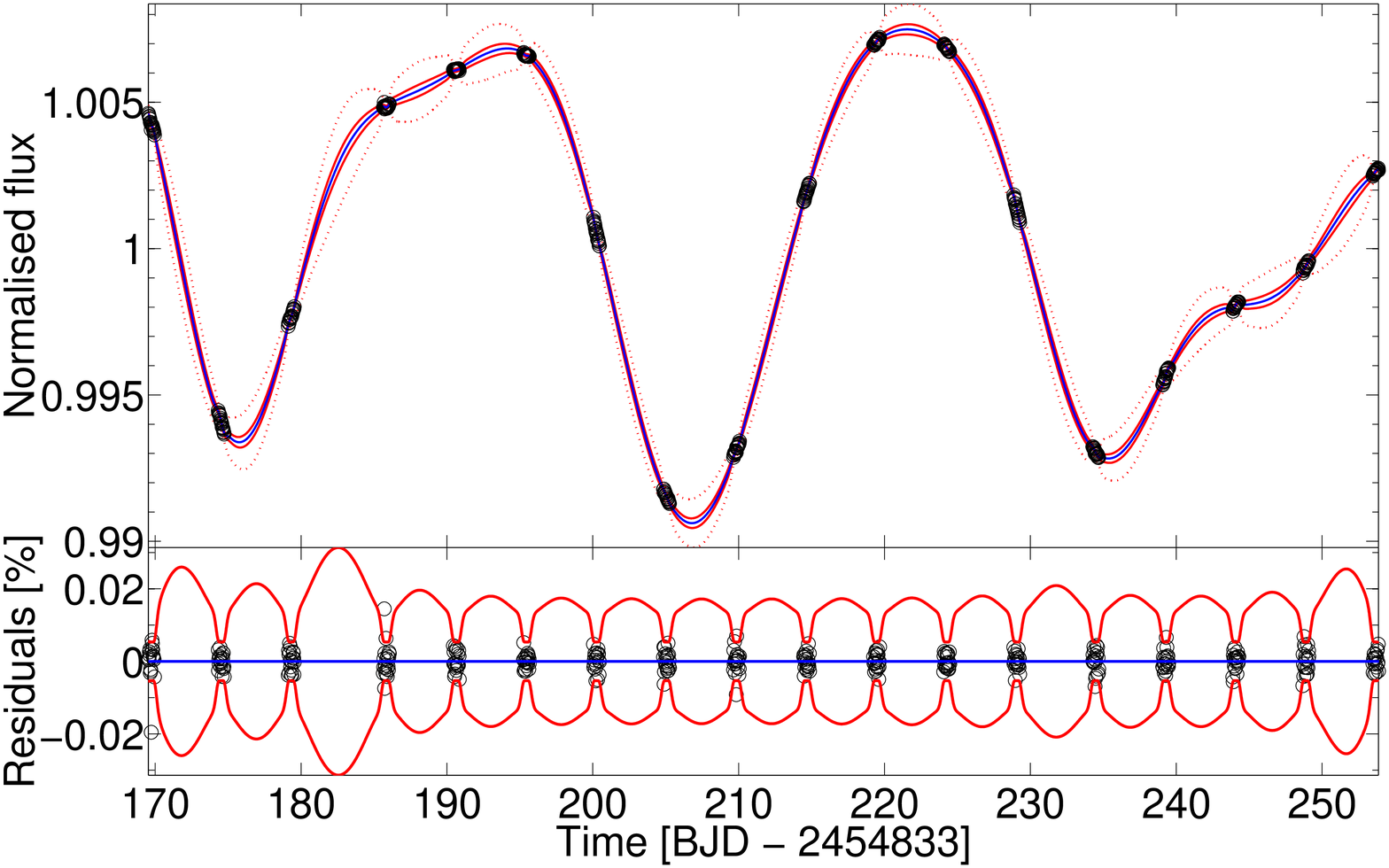,width=0.4\linewidth,clip=, trim = 10  0 80 20} \\
\epsfig{file=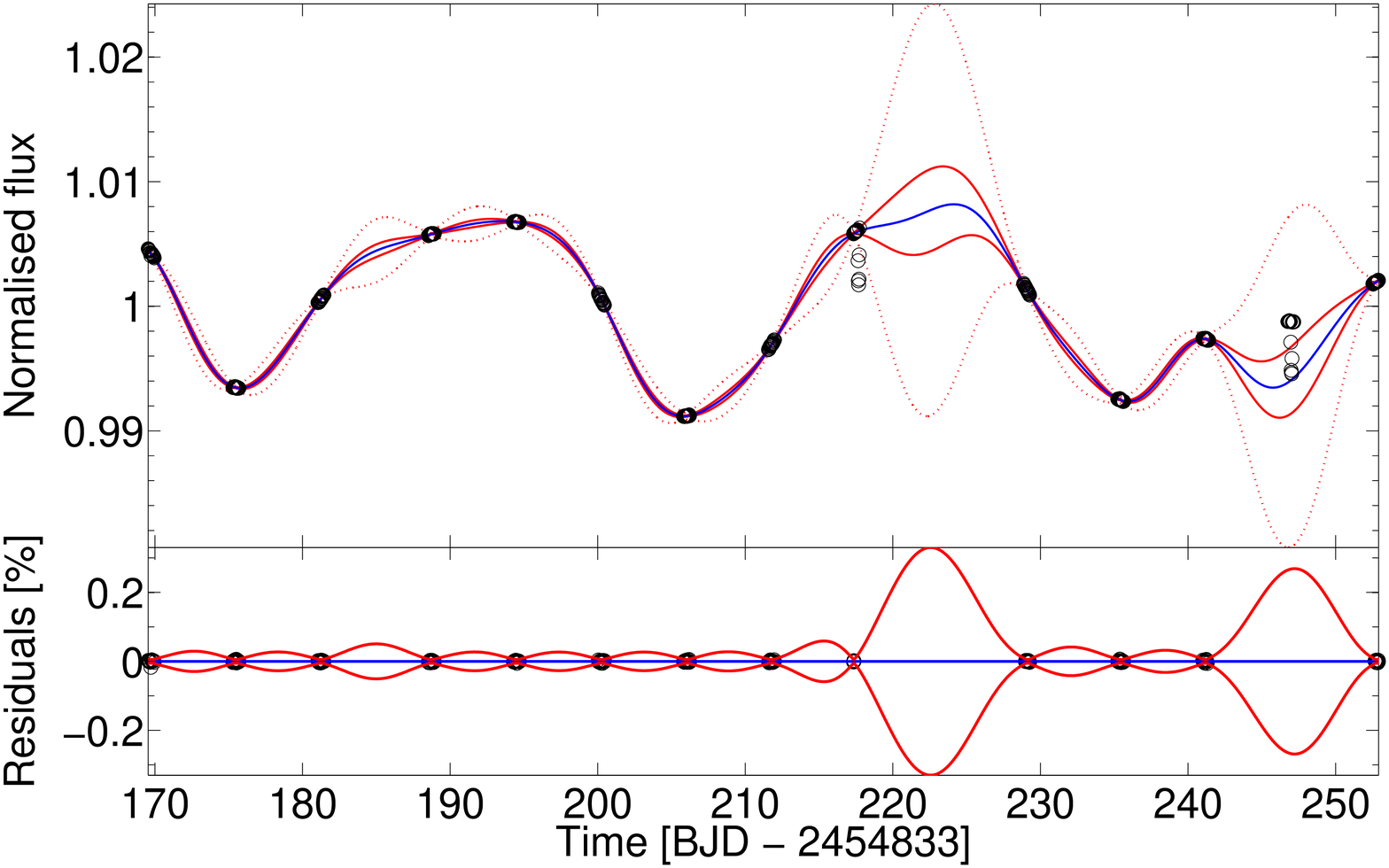,,width=0.4\linewidth,clip=, trim = 10 0 80 20} 
\epsfig{file=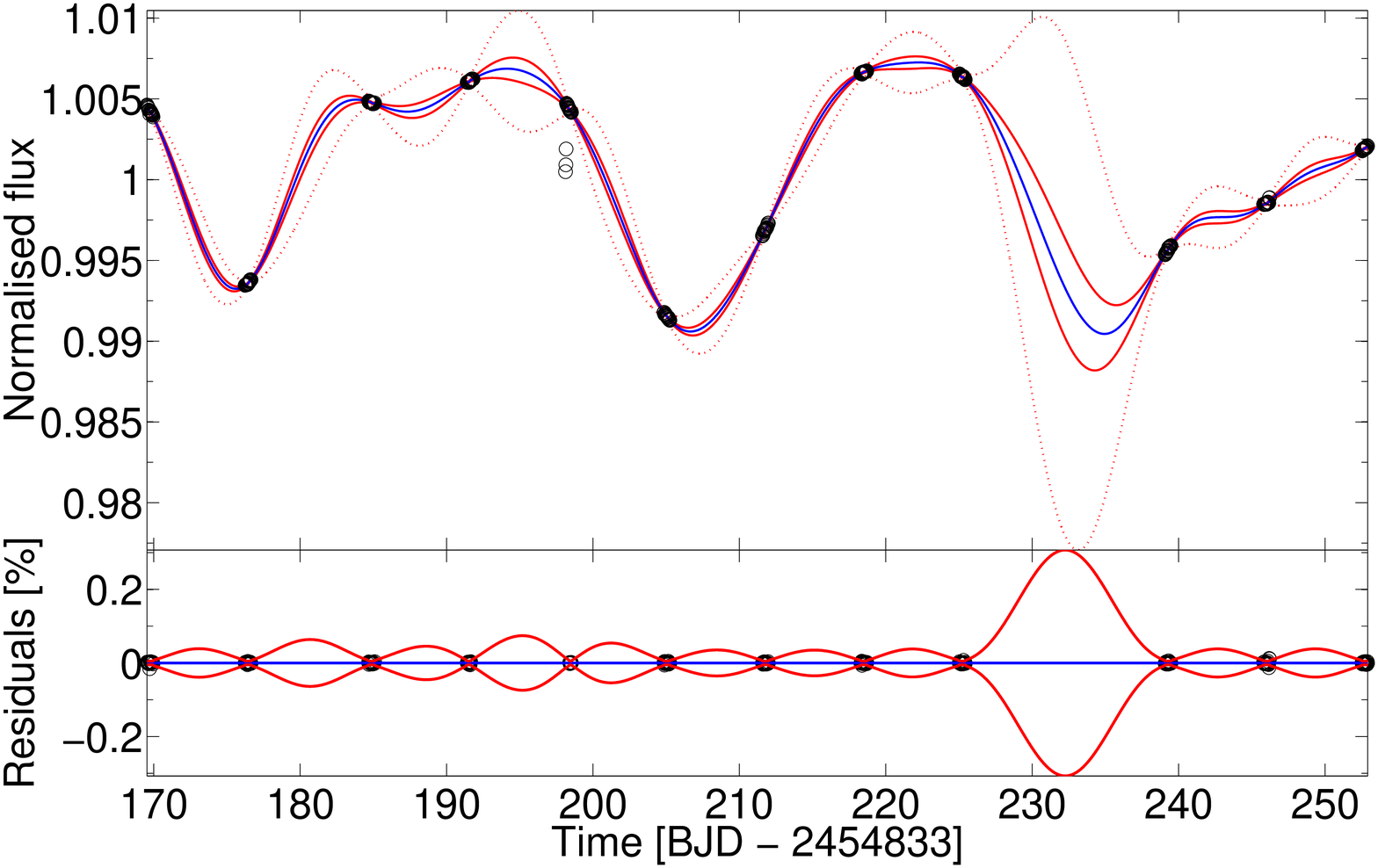,width=0.4\linewidth,clip=, trim = 10 0 80 20} \\
\end{tabular}
\caption{From left to right and from top to bottom: GPSC fit on: all the data of KIC 10748390 (Q-2), on a set of 10 yours of observation per day, every second day, every third day, 
every fourth day, every fifth day, every sixth day and every week. The data are plot in black, the blue line 
represents the  mean model, the red continuous and dotted line represent 1-$\sigma$ and 5-$\sigma$ uncertainty level. The plotted residuals are computed on the \textit{inliers} data.}
\label{fig:10748390}
\end{figure}

%% The \notetoeditor{TEXT} command allows the author to communicate
%% information to the copy editor.  This information will appear as a
%% footnote on the printed copy for the manuscript style file.  Nothing will
%% appear on the printed copy if the preprint or
%% preprint2 style files are used.

%% The eqnarray environment produces multi-line display math. The end of
%% each line is marked with a \\. Lines will be numbered unless the \\
%% is preceded by a \nonumber command.
%% Alignment points are marked by ampersands (&). There should be two
%% ampersands (&) per line.

%% If you wish to include an acknowledgments section in your paper,
%% separate it off from the body of the text using the \acknowledgments
%% command.

%% Included in this acknowledgments section are examples of the
%% AASTeX hypertext markup commands. Use \url without the optional [HREF]
%% argument when you want to print the url directly in the text. Otherwise,
%% use either \url or \anchor, with the HREF as the first argument and the
%% text to be printed in the second.

\acknowledgments
\textit{Acknowledgments}: We thank Ingo P. Waldmann  for his helpful suggestions and useful discussions.

%% To help institutions obtain information on the effectiveness of their
%% telescopes, the AAS Journals has created a group of keywords for telescope
%% facilities. A common set of keywords will make these types of searches
%% significantly easier and more accurate. In addition, they will also be
%% useful in linking papers together which utilize the same telescopes
%% within the framework of the National Virtual Observatory.
%% See the AASTeX Web site at http://www.journals.uchicago.edu/AAS/AASTeX
%% for information on obtaining the facility keywords.

%% After the acknowledgments section, use the following syntax and the
%% \facility{} macro to list the keywords of facilities used in the research
%% for the paper.  Each keyword will be checked against the master list during
%% copy editing.  Individual instruments or configurations can be provided 
%% in parentheses, after the keyword, but they will not be verified.

%{\it Facilities:} \facility{Nickel}, \facility{HST (STIS)}, \facility{CXO (ASIS)}.

%% Appendix material should be preceded with a single \appendix command.
%% There should be a \section command for each appendix. Mark appendix
%% subsections with the same markup you use in the main body of the paper.

%% Each Appendix (indicated with \section) will be lettered A, B, C, etc.
%% The equation counter will reset when it encounters the \appendix
%% command and will number appendix equations (A1), (A2), etc.

\appendix

%\section{Appendix material}

%% The reference list follows the main body and any appendices.
%% Use LaTeX's thebibliography environment to mark up your reference list.
%% Note \begin{thebibliography} is followed by an empty set of
%% curly braces.  If you forget this, LaTeX will generate the error
%% "Perhaps a missing \item?".
%%
%% thebibliography produces citations in the text using \bibitem-\cite
%% cross-referencing. Each reference is preceded by a
%% \bibitem command that defines in curly braces the KEY that corresponds
%% to the KEY in the \cite commands (see the first section above).
%% Make sure that you provide a unique KEY for every \bibitem or else the
%% paper will not LaTeX. The square brackets should contain
%% the citation text that LaTeX will insert in
%% place of the \cite commands.

%% We have used macros to produce journal name abbreviations.

%% AASTeX provides a number of these for the more frequently-cited journals.
%% See the Author Guide for a list of them.

%% Note that the style of the \bibitem labels (in []) is slightly
%% different from previous examples.  The natbib system solves a host
%% of citation expression problems, but it is necessary to clearly
%% delimit the year from the author name used in the citation.
%% See the natbib documentation for more details and options.

\clearpage

\clearpage

%% If you use the table environment, please indicate horizontal rules using
%% \tableline, not \hline.
%% Do not put multiple tabular environments within a single table.
%% The optional \label should appear inside the \caption command.

\clearpage

%% If the table is more than one page long, the width of the table can vary
%% from page to page when the default \tablewidth is used, as below.  The
%% individual table widths for each page will be written to the log file; a
%% maximum tablewidth for the table can be computed from these values.
%% The \tablewidth argument can then be reset and the file reprocessed, so
%% that the table is of uniform width throughout. Try getting the widths
%% from the log file and changing the \tablewidth parameter to see how
%% adjusting this value affects table formatting.

%% The \dataset{} macro has also been applied to a few of the objects to
%% show how many observations can be tagged in a table.

\clearpage

%% Tables may also be prepared as separate files. See the accompanying
%% sample file table.tex for an example of an external table file.
%% To include an external file in your main document, use the \input
%% command. Uncomment the line below to include table.tex in this
%% sample file. (Note that you will need to comment out the \documentclass,
%% \begin{document}, and \end{document} commands from table.tex if you want
%% to include it in this document.)

%% \input{table}

%% The following command ends your manuscript. LaTeX will ignore any text
%% that appears after it.

\end{document}

%% file: section_gp.tex
A Gaussian Process (GP) is a stochastic process commonly used for regression and classification tasks in fields of machine learning, statistics, time series analysis and geostatistics.
Intuitively, it can be viewed as a generalisation of a Gaussian probability distribution to {\em functions}, or alternatively as Gaussian distribution over functions.
Formally, a GP is a set of random variables, any finite set of which have a joint Gaussian distribution.
Here we adopt the notation of Rasmussen $\&$ Williams (2006).
We consider $\vec{x}$ as an input vector. 
A GP is completely specified by its mean function $m(\vec{x})$ and its covariance function $K(\vec{x_1},\vec{x_2})$, also often called the \emph{kernel} function.\\
For any pair of inputs $\vec{x_1}$ and $\vec{x_2}$, the mean and covariance functions define a joint distribution over function values:
$f(\vec{x_1}), f(\vec{x_2})$ $\sim ~\mathcal{N}(\vec{m},\vec{K})$, where $\mathcal{N}$ is a Gaussian distribution with mean vector $\vec{m}$ with elements $m_i = m(\vec{x}_i)$ and covariance matrix $\vec{K}$ with elements $K_{ij}=K(\vec{x}_i,\vec{x}_j)$.
In this sense a GP specifies a distribution over functions (the function-space), where
the mean  and  covariance are considered as priors on the function-space and are parametrized by so-called \textit{hyperparameters}.
%We used a constant $\vec{m}=\vec{0}$ or $\vec{m}=\vec{1}$ mean function throughout this work, but a general GP can use any mean function.
We used a constant $\vec{m}=\vec{1}$ mean function throughout this work, but a general GP can use any mean function.

Let us now consider a case where the observed function values $f(\vec{x})$ are contaminated with noise.
We obtain a finite number $i$ of noisy observed data $y_i$ of this function at a set of points $\vec{x}_i$, where $y_i =  f(\vec{x_i}) + n$ and  $n \sim \mathcal{N}[0,\sigma]$ is a Gaussian noise variable.
The goal is to reconstruct the values of the function $f(\vec{x})$.
Assuming we have a prior knowledge about the properties of the function we choose and specify the covariance $\vec{K}=K(\vec{X},\vec{X})$ and mean  $\vec{m} = m(\vec{X})$, where $\vec{X}=[{\vec{x}_1,\vec{x}_2,\vec{x}_2,...,\vec{x}_n}]^{\top}$ is a collection of input function arguments in a matrix form.
In the following we will skip writing the dependence on $\vec{X}$ for clarity.

\noindent Our prior on the function space has the form
\begin{equation}
p(\vec{f}~|~\vec{X}) =  \mathcal{N}\left( \vec{m}, \vec{K} \right)
\end{equation}
A set of observations $\vec{y}$ contains additive, white Gaussian noise with standard deviation $\sigma$:
\begin{equation}
p( \vec{y}~|~\vec{f} ) =  \mathcal{N} \left( \vec{f}, \sigma \vec{I} \right).
\end{equation}
where $\vec{I}$ is the identity matrix.
Integrating over $\vec{f}$ gives the marginal distribution on $\vec{y}$
\begin{equation}
p ( \vec{y} ) = \int p( \vec{y}~|~\vec{f})~ p( \vec{f} )~ d\vec{f} = \mathcal{N} [ \vec{y}~ |~ \vec{m}, \tilde{ \vec{K} }] ,
\end{equation}
where $\tilde{ \vec{K} }= \vec{K} +  \sigma^2 \vec{I}$. 

\noindent Now we would like to find the prediction for a test point $\vec{f_{*}}$ at locations of $\vec{X_{*}}$.
Joint probability of a set of observations $\vec{y}$ and test points $\vec{f_{*}}$ is
\begin{equation}
\label{eqn:gp_marginal}
p 
\left(
\left[
\begin{matrix}
\vec{y    } \\
\vec{f_{*} }
\end{matrix}
\right]
\right)  
% ////////////
=
\mathcal{N}
\left(
\left[ 
\begin{matrix}
\vec{m} \\
\vec{m}_{*}
\end{matrix}
\right]
,
\left[ 
\begin{matrix}
\tilde{\vec{K}} & \vec{k} \\
\vec{k}^{\top} & \vec{c} \\
\end{matrix}
\right]
\right)
% | \vec{f}) p(\vec{f}) d\vec{f} = \mathcal{N}\left[ \vec{t} | \vec{0}, \vec{C} \right]  
\end{equation}
where $\vec{m}_{*} = m(\vec{X}_{*})$ and $\vec{k} = K(\vec{X},\vec{X_{*}})$, $\vec{c}=K(\vec{X}_{*},\vec{X_{*}})$. 

\noindent Hence, conditioning on observed values of $\vec{y}$ allows us to calculate the conditional probability for $\vec{f}_{*}$, which will be a Gaussian distribution:  
\begin{equation}
\label{eqn:gp_conditional}
p(\vec{f}_{*}~|~\vec{y},\vec{X},\vec{X_{*}}) = \mathcal{N}(\vec{m}_{*}+\vec{k}^{\top} \tilde{\vec{K}}^{-1} (\vec{y} - \vec{m}) \ , \ \vec{c} - \vec{k}^{\top} \tilde{\vec{K}}^{-1} \vec{k}).
\end{equation}
This is the mean and the variance for the function $\vec{f}_{*}$ at a test point $\vec{x}_{*}$.
The variance on this value will, in principle, depend on the noise variance $\sigma^2$ (which is assumed known), covariance function properties and proximity of the points in the training set. 
Figure \ref{fig:gpdemo} demonstrates GP regression, where the true underlying function (white dashed) is observed at 30 irregularly spaced points (white dots) and these observations are affected by significant amount of noise. 
Posterior GP (black solid) is conditioned on these observations and is using a prior in a form of stationary covariance function and zero-mean function. 
Intensity map shows a normalised probability on the function space and black dotted lines correspond to 1-$\sigma$ uncertainty on the function value.

Making a prediction using a GP requires calculating the kernel function between the test and training data. 
This involves an inverse of the kernel matrix and further linear algebra operations, making this process relatively straightforward, although relatively computationally expensive, as it scales as $\mathcal{O}(N^3)$ in the number of training points.

Gaussian Process for regression does not assume any particular functional form for the function $\vec{f}$, and in this sense it is a non-parametric technique.
It does, however, use hyperparameters which describe the statistical properties of $\vec{f}$. 
The functional form for the covariance is chosen to match the prior beliefs about the statistical properties of the function $\vec{f}$. 
When there is no theoretical motivation for any particular choice of the covariance function form and values its hyperparameters, the functional form is chosen from a popular set of functions used in the community, for example Radial Basis Function (RBF), polynomial kernel, rational quadratic, etc.
The hyperparameters are then optimised, often using cross-validation of the training data. 

In this work we used a combination of RBF kernels defined as
\begin{equation}
\label{eqn:rbf_kernel}
	K(\vec{x}_1,\vec{x}_2) = \sum \limits_{i=1}^2 \lambda_i \exp \left(  -\frac{1}{2 l_i^2} (\vec{x}_1-\vec{x}_2)^{\top}(\vec{x}_1-\vec{x}_2) \right)
\end{equation}
where $\lambda$ and $l$ are kernel hyperparameters. %We used a mean prior of 1.
The choice of this kernel and the optimization of the hyperparameters are described in the following Section \ref{sec:model}.

%% file: GP_final.bbl
\begin{thebibliography}{unsrtnat}



\bibitem[Ballard et al.(2011)]{2011ApJ...743..200B} Ballard, S., Fabrycky, 
D., Fressin, F., et al.\ 2011, \apj, 743, 200 

\bibitem[Barclay et al.(2013)]{2013Natur.494..452B} Barclay, T., Rowe, 
J.~F., Lissauer, J.~J., et al.\ 2013, \nat, 494, 452 

\bibitem[Basri et al.(2011)]{2011AJ....141...20B} Basri, G., Walkowicz, 
L.~M., Batalha, N., et al.\ 2011, \aj, 141, 20 

\bibitem[Basri et al.(2013)]{2013ApJ...769...37B} Basri, G., Walkowicz, 
L.~M., \& Reiners, A.\ 2013, \apj, 769, 37

\bibitem[Borucki et al.(2010)]{2010Sci...327..977B} Borucki, W.~J., Koch, 
D., Basri, G., et al.\ 2010, Science, 327, 977 

\bibitem[Borucki et al.(2011)]{2011ApJ...736...19B} Borucki, W.~J., Koch, 
D.~G., Basri, G., et al.\ 2011, \apj, 736, 19 

\bibitem[Borucki et al.(2013)]{Borucki13} Borucki, W.~J., Agol, 
E., Fressin, F., et al.\ 2013, DOI:   10.1126/science.1234702

\bibitem[Chapman et al.(2013)]{2013MNRAS.429..165C} Chapman, E., Abdalla, 
F.~B., Bobin, J., et al.\ 2013, \mnras, 429, 165 

\bibitem[Csizmadia et 
al.(2013)]{2013A&A...549A...9C} Csizmadia, S., Pasternacki, T., Dreyer, C., et al.\ 2013, \aap, 549, A9

\bibitem[J. L. Christiansen, et al. (2012)]{Christiansen12} Christiansen, J. L., et al.\ 2012, Kepler Data Characteristics Handbook (KSCI-19040-003)

\bibitem[Ford et al.(2012)]{2012ApJ...750..113F} Ford, E.~B., Fabrycky, 
D.~C., Steffen, J.~H., et al.\ 2012, \apj, 750, 113 

\bibitem[Garc{\'{\i}}a et al.(2011)]{2011MNRAS.414L...6G} Garc{\'{\i}}a, 
R.~A., Hekker, S., Stello, D., et al.\ 2011, \mnras, 414, L6 

\bibitem[Gibson et al.(2012)]{2012MNRAS.419.2683G} Gibson, N.~P., Aigrain, 
S., Roberts, S., et al.\ 2012, \mnras, 419, 2683 

\bibitem[Gilliland et al.(2010)]{2010ApJ...713L.160G} Gilliland, R.~L., 
Jenkins, J.~M., Borucki, W.~J., et al.\ 2010, \apjl, 713, L160

\bibitem[Haario (2006)]{Haario2006} Haario, H., Laine, L., Mira, A., Saksman, E., 2006,
Statistics and Computing, 16, 339

\bibitem[Howard et al.(2013)]{2013arXiv1310.7988H} Howard, A.~W., 
Sanchis-Ojeda, R., Marcy, G.~W., et al.\ 2013, arXiv:1310.7988

\bibitem[Jenkins et al.(2010)]{2010ApJ...713L..87J} Jenkins, J.~M., 
Caldwell, D.~A., Chandrasekaran, H., et al.\ 2010a \apjl, 713, L87 

\bibitem[Jenkins et al.(2010)]{2010ApJ...713L.120J} Jenkins, J.~M., 
Caldwell, D.~A., Chandrasekaran, H., et al.\ 2010b, \apjl, 713, L120 

\bibitem[Lanza et 
al.(2010)]{2010A&A...520A..53L} Lanza, A.~F., Bonomo, A.~S., Moutou, C., et al.\ 2010, \aap, 520, A53 

\bibitem[Mandel 
\& Agol(2002)]{2002ApJ...580L.171M} Mandel, K., \& Agol, E.\ 2002, \apjl, 580, L171 

\bibitem[Murphy(2012)]{2012MNRAS.422..665M} Murphy, S.~J.\ 2012, \mnras, 
422, 665 

\bibitem[Oshagh et 
al.(2013)]{2013A&A...556A..19O} Oshagh, M., Santos, N.~C., Boisse, I., et al.\ 2013, \aap, 556, A19

\bibitem[Pepe et al.(2013)]{2013arXiv1310.7987P} Pepe, F., Collier Cameron, 
A., Latham, D.~W., et al.\ 2013, arXiv:1310.7987 

\bibitem[Quinonero Candela and Rassmussen (2005)]{QuinoneroCandelaRassmussen2005}
  Qui\~{n}onero,  Candela and Rasmussen
\newblock {\em A unifying view of sparse approximate Gaussian process regression.}
\newblock Journal of Machine Learning Research, 2005


\bibitem[Rasmussen and William (2006)]{RasmussenWilliam2006} Rasmussen, C.~ and Williams, C.~.
\newblock {\em Gaussian Processes for Machine Learning}.
\newblock MIT Press, 2006.

\bibitem[Sanchis-Ojeda et al.(2013)]{2013ApJ...774...54S} Sanchis-Ojeda, 
R., Rappaport, S., Winn, J.~N., et al.\ 2013, \apj, 774, 54 

\bibitem[Seikel et al.(2012)]{2012JCAP...06..036S} Seikel, M., Clarkson, 
C., \& Smith, M.\ 2012, \jcap, 6, 36 

\bibitem[Smith et al.(2012)]{2012PASP..124.1000S} Smith, J.~C., Stumpe, 
M.~C., Van Cleve, J.~E., et al.\ 2012, \pasp, 124, 1000 

\bibitem[Steffen et al.(2013)]{2013MNRAS.428.1077S} Steffen, J.~H., 
Fabrycky, D.~C., Agol, E., et al.\ 2013, \mnras, 428, 1077 

\bibitem[Thatte et 
al.(2010)]{2010A&A...523A..35T} Thatte, A., Deroo, P., \& Swain, M.~R.\ 2010, \aap, 523, A35
 
 \bibitem[J. E. Van Cleve & D. A. Caldwell (1000)]{VanCleve00}Van Cleve, J. E., \& Caldwell,D. A., Kepler Instrument Handbook (KSCI-19033)
 
\bibitem[Waldmann(2012)]{2012ApJ...747...12W} Waldmann, I.~P.\ 2012, \apj, 
747, 12

\bibitem[Waldmann(2013)]{Waldmann2013} Waldmann, I.~P.\ 2013, arXiv:1302.6714

\bibitem[Waldmann et al.(2013)]{2013ApJ...766....7W} Waldmann, I.~P., 
Tinetti, G., Deroo, P., et al.\ 2013, \apj, 766, 7 




\end{thebibliography}
